\documentclass{iopart}
\usepackage{iopams}
  \expandafter\let\csname equation*\endcsname\relax
  \expandafter\let\csname e\endcsname\relax
  \expandafter\let\csname endequation*\endcsname\relax
\usepackage{amsfonts,amssymb,amsmath,amsthm}
\usepackage{nomencl}
\usepackage[utf8]{inputenc}
\usepackage{lmodern}
\usepackage[T1]{fontenc}
\usepackage{graphicx}
\usepackage{batex}
\usepackage[normalem]{ulem}
\usepackage{microtype}
\usepackage{hyperref}
\usepackage{fancyhdr}
\newcommand{\eq}[1]{ Eq.~(\ref{eq:#1}) }

\newcommand{\pspace}{\ensuremath{\Gamma} } 
\newcommand{\ospace}{\ensuremath{\Omega} } 
\newcommand{\cell}{\ensuremath{\mathcal{C}} } 
\newcommand{\vcell}{\ensuremath{\Pi} } 

\newcommand{\traj}[1]{\ensuremath{\uline{#1}} } 
\newcommand{\micden}{\ensuremath{\varrho} } 
\newcommand{\mesden}{\ensuremath{\rho} } 
\newcommand{\trav}[1]{\ensuremath{\left\langle\!\left\langle #1 \right\rangle\!\right\rangle}}
\newcommand{\eav}[1]{\ensuremath{\left\langle #1 \right\rangle} }
\newcommand{\mvert}{\,\vert\,} 
\newcommand{\mVert}{\,\Vert\,} 

\newcommand{\tobs}{\ensuremath{\tau_{\mathrm{obs}}} } 
\newcommand{\tmic}{\ensuremath{\tau_{\mathrm{mic}}} } 
\newcommand{\tmom}{\ensuremath{\tau_{\pvec p}} } 
\newcommand{\tcoord}{\ensuremath{\tau_{\pvec q}} } 
\newcommand{\tconf}{\ensuremath{\tau_{\mathrm{conf}}} } 

\newcommand{\pvec}[1]{\ensuremath{\boldsymbol{#1}} } 
\newcommand{\obs}{\ensuremath{A} } 
\newcommand{\meas}{\ensuremath{\mathcal M} } 

\newcommand{\sent}[1]{\ensuremath{\mathcal S}\left[#1\right] } 
\newcommand{\kldiv}[1]{\ensuremath{\mathcal D_{\mathrm{KL}}}\left[#1\right] } 

\newcommand{\intrv}{\ensuremath{s_{\mathrm{int}}} } 
\newcommand{\visrv}{\ensuremath{s_{\mathrm{vis}}} } 
\newcommand{\jumprv}{\ensuremath{s_{\mathrm{mot}}} } 
\newcommand{\sysrv}{\ensuremath{s_{\mathrm{sys}}} } 
\newcommand{\medrv}{\ensuremath{s_{\mathrm{med}}} } 
\newcommand{\totrv}{\ensuremath{s_{\mathrm{tot}}} } 
\newcommand{\cgrv}{\ensuremath{s_{\mathrm{cg}}} } 

\newcommand{\visent}{\ensuremath{S_{\mathrm{vis}}} } 
\newcommand{\intent}{\ensuremath{S_{\mathrm{int}}} } 
\newcommand{\jumpent}{\ensuremath{S_{\mathrm{mot}}} } 

\newcommand{\sysent}{\ensuremath{S_{\mathrm{sys}}} } 
\newcommand{\medent}{\ensuremath{S_{\mathrm{med}}} } 
\newcommand{\totent}{\ensuremath{S_{\mathrm{tot}}} } 

\newcommand{\fgent}{\ensuremath{S_{\mathrm{fg}}} } 
\newcommand{\cgent}{\ensuremath{S_{\mathrm{cg}}} } 
\newcommand{\klent}{\ensuremath{S_{\mathrm{rel}}} } 

\newcommand{\visep}{\ensuremath{\Sigma_{\mathrm{vis}}} } 
\newcommand{\jumpep}{\ensuremath{\Sigma_{\mathrm{mot}}} } 
\newcommand{\totep}{\ensuremath{\Sigma_{\mathrm{tot}}} } 

\newcommand{\viseprv}{\ensuremath{\sigma_{\mathrm{vis}}} } 
\newcommand{\inteprv}{\ensuremath{\sigma_{\mathrm{int}}} } 
\newcommand{\jumpeprv}{\ensuremath{\sigma_{\mathrm{mot}}} } 
\newcommand{\medeprv}{\ensuremath{\sigma_{\mathrm{med}}} } 
\newcommand{\toteprv}{\ensuremath{\sigma_{\mathrm{tot}}} } 

\newcommand{\nom}{\nomenclature}

\begin{document}
\article{PREPRINT}{} 

\title{Stochastic Thermodynamics, Reversible Dynamical Systems and Information Theory}

\author{Bernhard Altaner}
\address{Max Planck Institute for Dynamics and Self-Organization, G\"ottingen, Germany} 
\ead{bernhard.altaner@ds.mpg.de}

\begin{abstract}
  \textit{Dated: \today}\\
  Stochastic Thermodynamics (ST) extends the notions of classical thermodynamics to trajectories taken from a nonequilibrium ensemble.
This extension yields a simple approach to fluctuation relations in small systems.
Multiple time- and length scales play an important role for measurements but also for the foundations of nonequilibrium statistical mechanics.
Here, under the assumptions of local equilibrium we derive the trajectory functionals of ST in the context of reversible deterministic thermostats.
Further, the connection to previous work is made.

\noindent{\it Keywords\/}: Stochastic thermodynamics, entropy, information theory, reversible thermostats, entropy production.
\end{abstract}
\pacs{
05.70.-a: Entropy thermodynamics;
89.70.Cf: Entropy in information theory;
05.70.Ln: Nonequilibrium and irreversible thermodynamics;
05.40.-a: Fluctuation phenomena, random processes, noise, and Brownian motion.
 }

\maketitle

%
%
%
%
%

\tableofcontents
\clearpage

\section{Introduction}
Arguably the most important open issue in statistical mechanics is the quest to understand the role of entropy and entropy production in systems far from equilibrium.
In equilibrium situations, diffusion usually smoothens out gradients such that one would not expect interesting structure.
Out of equilibrium, the situation is different and it is usually crucial to consider (dynamics on) multiple scales.
This goes beyond the classical understanding of entropy as a thermodynamic state variable.
%

Since computer simulations have become feasible, scientists have been trying to model and understand the microscopic dynamics of systems in contact with thermodynamic environments.
In such simulations the environment is usually modelled by modified equations of motion, so-called thermostats \cite{Jepps+Rondoni2010}.
Thermostats can usually be divided into deterministic and stochastic thermostats.

Soon after their introduction, dynamical systems theory got interested in deterministic thermostats.
Phase space contraction in such systems was understood to be related to physical entropy production.
Also, the first fluctuation relations for nonequilibrium systems arose in that context \cite{Evans_etal1993,Gallavotti+Cohen1995}.

Stochastic thermostats have a longer history and date back to the works of Einstein, Smoluchowski and Langevin in the early 20th century.
Hill \cite{Hill1977} and later Schnakenberg \cite{Schnakenberg1976} connected stochastic dynamics with dissipation in small systems.
Since the late 90s, fluctuation relations have also been found for stochastic dynamics \cite{Kurchan1998,Lebowitz+Spohn1999,Maes2004,Seifert2005}.

Fluctuation relations are concerned with the probability of rare events that seem to act contrary the 2nd law of thermodynamics.
Whereas this is a phenomenon that cannot be observed in the macroscopic world, such fluctuations are quite common in the microscopic world of macromolecules in solution.
The latter have become the paradigm of what today is called \emph{stochastic thermodynamics}.
Because of the advent of modern measurement and simulation techniques, much research is done on this interface between physics, chemistry and biology.

For instance, the work relations by Jarzynski and Crooks, which allowed an experimental measurement of the free energy landscape of macromolecules, have been of huge impact \cite{Jarzynski1997, Crooks1999}.
While the former was originally derived in the context of dynamical systems, the latter was found for stochastic dynamics.
Today, they are both understood as consequences of the above mentioned fluctuation theorems.

Besides others \cite{Verlinde2011}, this emergence is just one hint of a deeper connection of the entropy concepts in dynamical systems theory and stochastic dynamics.
The reason for this is the intimate connection of entropy with Shannon's information theory \cite{Shannon1948,Jaynes1957,Sagawa+Ueda2010,Mandal+Jarzynski2012}.
However, this is not generally acknowledged and entropy, but even more so, entropy production often seem to be an almost emotional topic with many different points of view.
Such views range from restricting entropy to the realm of classical thermodynamics to promoting variation principles for entropy production as the general mechanism for nonequilibrium steady states.
Some people only allow the phase space contraction in thermostatted dynamics to be connected with dissipation, other critize the phenomenological equations of motion as having no connection to the true microscopic dynamics.

In this paper, we follow earlier works \cite{Breymann_etal1998,Vollmer2002,Maes+Netocny2003,Maes2004,Gaspard2004} and try to highlight the connections of dynamical systems, statistical mechanics and information theory.
In order to connect the fields beyond pure mathematical considerations, we have to discuss some subtle questions:
\begin{itemize}
  \item How do the crucial assumptions of (stochastic) thermodynamics rely on a notion of separation of time- and length-scales?
  \item How is the information which is contained in a measurement affected by these scales?
  \item How does thermodynamic entropy arise in the context of information theory?
\end{itemize}
Finally, this will enable us to derive the modern notion of stochastic thermodynamics \cite{Seifert2012}, which is based on entropic trajectory functionals that can be obtained from deterministic equations of motion.
Throughout the paper we will motivate interpretations directly at the level of definitions, to avoid possibly misleading a-posteriori interpretations that could lead to inconsistencies.

The work is organized as follows:
Section \ref{sec:microscopic_basics} reviews basic notions of dynamical systems theory with a focus on Hamiltonian systems.
In section \ref{sec:stochastic_thermodynamics} we introduce stochastic equations of motion and briefly review Seifert's approach to stochastic thermodynamics \cite{Seifert2012}.
Section \ref{sec:thermostats} motivated variants of thermostatted equations of motion to provide a framework for further discussions \cite{Samoletov_etal2007}.
The main part of the work is section \ref{sec:information_entropy}.
There, we formulate the crucial assumptions of stochastic thermodynamics in the context of determinisic dynamics.
Under these assumptions, we are able to construct exact correspondences of the stochastic entropy functionals.
This is achieved by applying Shannon's notion of information to two different densities, reflecting the (unobservable) microscopic probability densities and the (observable) measured densities, respectively.

\section{Microscopic Basics}
\label{sec:microscopic_basics}

\subsection{Hamiltonian dynamics}

We will start with deterministic Hamiltonian dynamics on a phase space $\pspace$.
A point $x$ \nom{$x$}{point in phase space (microstate)}$ = \left( \pvec q, \pvec p \right) \in \pspace$ represents the state of a \nom{$N$}{number of particles} $N$-particle system in \nom{$d$}{spatial dimension of physical space}$d$-dimensional physical space consisting of the coordinates \nom{$\pvec q$}{(generalized) coordinates} $\pvec q$ and \nom{$\pvec p$}{(generalized) momenta} $\pvec p$ of all particles.%
\footnote{
To denote vectors in physical space, we will use bold symbols like $\pvec q$.
Elements of more general multi-dimensional sets will not receive any special decoration.
}
For brevity, here and in the following we use the short notation $\pvec q = \set{\pvec q_k }_{k=1}^N$, $\pvec p = \set{ \pvec p_k }_{k=1}^N$ when no ambiguity can arise.
The equations of motion
\begin{equation}
  \dot{ \pvec{ q}} = \frac{\del H}{\del \pvec p} = \frac{\pvec p}{m},\qquad
  \dot{ \pvec{ p}} = -\frac{\del H}{\del \pvec q} = -\del_{\pvec q}V(\pvec q),
  \label{eq:HamiltonianEOM}
\end{equation}
are governed by the Hamiltonian \nom{$H(\pvec q, \pvec p)$}{Hamiltonian}
\begin{equation}
  H(x) =  V(\pvec q)+\sum \frac{\pvec {p ^2}}{2m}.
  \label{eq:Hamiltonian}
\end{equation}
In our notation the term $\sum \pvec p^2 /2m$ \nom{$m$}{mass of a particle} in \eq{Hamiltonian} is short for $\sum_{k=1}^N \frac{\pvec{p}_k^2}{2 m_k}$ including the (possibly different) masses $m_k$.

For macroscopic physical systems the number of particles, $N\sim \bigo{10^{23}}$, is very large.
Even for much smaller, mesoscopic%
\footnote{
By mesoscopic we mean the range of scales between typical molecular scales ($1\text{\AA} =10^{-10}$m) to typical macroscopic scales ($10^{-3}$ -- $10^{0}$m)
}
systems this number is still large, mostly because the system of interest is not placed in vacuum but in an environment consisting of many particles.

This renders the microscopic equations of motion (\ref{eq:HamiltonianEOM}) unfeasible for simulating meso- or macroscopic systems in thermodynamic environments.
Even with state-of-the-art supercomputers, simulations of no more than a few \mbox{($10^4$--$10^6$)} particles on small time scales ($10^2$--$10^4$ns) are possible.
Hence, finding effective dynamical models for fewer degrees of freedom is one of the main interests of modern (statistical) physics.

Another problem  with using the equations (\ref{eq:HamiltonianEOM}) is that they need to be augmented with microscopic initial conditions.
Such detailed information is never accurately available for real complex systems.
Together with numerical inaccuracies, this puts a perspective on the value of a single \emph{microscopic} trajectory.

To overcome these problems, statistical mechanics introduces the notion of \emph{ensembles}.
Mathematically, an ensemble is a probability measure on phase space $\pspace$ \nom{\pspace}{space of microscopic configurations (phase space)}.
Throughout this work we assume that an ensemble can be specified by a (possibly time-dependent) phase space density $\micden^{(t)}(x)$.%
\nom{\micden}{microscopic density}%
\footnote{
This excludes, for instance, the so-called SRB-measures \cite{Ruelle1999} that appear asymptotically in the context of non-Hamiltonian dynamics, as described below.}
Ensembles are usually taken to reflect the probability of finding a certain microstate when only a macro- or mesoscopic state can be specified due to the finite measurement resolution.

Because of the lack of experimental accessibility, one is not interested in the microscopic state $x$ but rather in the current value of some measurable observable \obs \nom{\obs}{phase space function (observable)}.
In classical thermodynamics, these observables where usually macroscopic (bulk) properties of largely homogeneous systems.
Nowadays, due to a vast improvement in measurement techniques, the measurements are taken on \emph{mesoscopic} scales that lie in between the molecular (microscopic) and the bulk (macroscopic) scale.
The qualitative difference between meso- and macroscopic scales is that on the former fluctuations can be observed and may play an important role, whereas for the latter they can be neglected.
In what follows, we will always think of an observable being defined on mesoscopic scales though the discussion remains valid for macroscopic observables.

Mathematically, an observable is a mapping
\begin{equation}
  \begin{split}
    \obs: \pspace &\to \mathbb R,\\
  x &\mapsto \obs(x).
  \end{split}
  \label{eq:DefinitionObservable}
\end{equation}
A measurement \meas \nom{\meas}{measurement} consists of measuring an number of observables $\obs_i$.
We summarize the outcome of that measurement in an abstract (possibly multi-dimensional) mesoscopic observable $\omega$ \nom{$\omega$}{mesoscopic state (outcome of a measurement)}.
The measurement \meas is a surjective mapping from phase space $\pspace$ to an abstract space of observations, $\ospace$: \nom{\ospace}{space of observations}

\begin{equation}
  \begin{split}
    \meas: \pspace &\to \ospace,\\
  x &\mapsto \omega(x).
  \end{split}
  \label{eq:DefinitionMeasurement}
\end{equation}

The \emph{ensemble average} \nom{$\eav{\cdot}^{\left( \tau \right)}$}{ensemble average (at time $\tau$)} of an observable  $\obs$ is
\begin{equation}
  \eav{ \obs}^{(t)} = \int_\pspace \micden^{(t)}(x) \obs(x) \df x.
  \label{eq:EnsembleAverage}
\end{equation}

An empirical sampling of the distribution of the values of $\obs$ is done by measuring this observable for many experiments or simulations.

For real experiments, already the single measurement of an observable implicitly features a time average over some finite observation time $\tobs$: \nom{$\tobs$}{time-scale of observation}:
\begin{equation}
  \bar{\obs_{\tobs}}(t) = \frac{1}{\tobs}\int_{t}^{t+\tobs} \obs\left( x\left( t \right) \right) \df t.
  \label{eq:TimeAverage}
\end{equation}
\nom{$\bar{\cdot}(\tau)$}{time average (at time $\tau$)}%
One usually assumes that there is a separation of time scales $\tau_{\obs} \sim \tobs \gg \tmic$ \nom{\tmic}{typical microscopic time-scale} between the time-scales $\tau_\obs$ of the evolution of the observable and an  unobservable microscopic time scale $\tmic$ associated with the dynamics in phase space.

Given that, one can choose $\tobs$ such that $\bar \obs(t)$ is effectively independent of this scale.

\subsection{Ergodicity and time-sampling of phase space}
\emph{Equilibrium thermodynamics} relies on ergodicity, which is a crucial assumption relating time- and ensemble averages.
The ergodic hypothesis assumes that there exists a stationary measure $\micden^{\infty}$ \nom{$\micden^\infty$}{stationary phase space density} for physical many-particle systems such that
\begin{equation}
  \lim_{\tobs \to \infty}  \frac{1}{\tobs}\int_{t}^{t+\tobs} \obs\left( x\left( t \right) \right) \df t =: \int_\pspace \micden^{\infty}(x) \obs(x) \df x
  \label{eq:TimeAverageInfinity}
\end{equation}
exists for any observable $\obs$  and is independent of the time $t$ when one starts the time-average (no aging).
In other words, the ergodic hypothesis states that there exists an ensemble that samples phase space in the same way as an infinitely long trajectory would do asymptotically.
Proving the ergodic hypothesis for actual systems is usually very difficult.
However, it constitutes the dynamical microscopical approach to equilibrium thermodynamics combining statistical mechanics with dynamical systems theory.


\subsection{Equilibrium and nonequilibrium ensembles}

In contrast to the situation above, modern statistical mechanics seeks to understand the foundations of nonequilibrium systems.
Physical nonequilibrium systems are always in a time-dependent transient state which may or may not relax towards an equilibrium on accessible time scales.
Therefore for nonequlibrium situations it is not enough to consider an asymptotic density $\micden^{\infty}$.

The dynamical behaviour of the density $\micden^{(t)}$ is governed by the flow \nom{$\Phi$}{evolution operator (flow)}
\begin{equation}
  \begin{split}
    \Phi^{(\tau)}: \pspace \times \mathbb R &\to \pspace,\\
    x(t) &\mapsto x(t+\tau)
  \end{split}
  \label{eq:DefinitionFlow}
\end{equation}
generated by \eq{HamiltonianEOM}.

If the flow of a Hamiltonian system on the hypersurface defined by $H(x) \stackrel{!}{=} E_0 = H(x(0))$ is ergodic, it always yields the so-called \emph{microcanonical distribution} for constant energy $E_0$,
\begin{equation}
  \micden^{\infty}(x) \propto \chi_{H(x) = E_0},
  \label{eq:MicrocanonicalDistribution}
\end{equation}
where $\chi$ \nom{$\chi_\cell(x)$}{characteristic function of set $\cell$} denotes the characteristic function.


However, in experiments one specifies the temperature $T$ \nom{$T$}{temperature} of the environment (heat bath) rather than the total energy.
Also, physical observables should depend only on the state of the system and not on the state of its environment.
From classical statistical physics we would expect a (Maxwell-)Boltzmann distribution for the degrees of freedom of the system.
One is inclined to ask for the ``right'' reduced equations of motion in a reduced phase space $\pspace_{sys}$ consisting of the degrees of freedom of the system only.
Because we ignore the details of the interaction with the environment, a trajectory produced by such reduced dynamics does certainly not describe the actual microscopic motion.
However, as we argued above, single trajectories or instantaneous configuration do not really matter.
The important thing is that the new dynamics generates trajectories that sample the phase space in accordance with macroscopic constraints.
In other words, $\micden^{\infty}$ must be the (equilibrium) distribution expected from the ensemble belonging to the macroscopic parameters of the environment (like temperature, pressure etc.).%
\footnote{That this trajectory also samples the transient regime equally well is usually tacitly assumed, but has no deeper justification.}

But since Hamiltonian dynamics will only generate microcanonical distributions, the reduced dynamics on $\pspace_{sys}$ will be in general non-Hamiltonian.
In the following two sections we review two different approaches to such reduced dynamics.


\section{Stochastic Thermodynamics}

\label{sec:stochastic_thermodynamics}

Stochastic thermodynamics is a modern paradigm for the treatment of (small) systems in a thermodynamic environment.
It is based on the notion of stochastic differential equations (Langevin equations) or the corresponding Fokker-Planck or path integral descriptions \cite{Gardiner2004,Schulman2005}.
Physically, the stochastic term that appears in the equations of motion models the interaction of the heat bath with the system.
In the following paragraphs we review the basic concepts of stochastic thermodynamics.
The presentations closely follows Seifert's work \cite{Seifert2005,Seifert2011} (for an extensive review see Ref.~\cite{Seifert2012}) though notation may vary.

\subsection{Langevin equation}
The stochastic evolution of the system on the reduced phase space $\pspace_{\mathrm{sys}}$ can be described by a Langevin equation:
  \begin{subequations}
\begin{align}
    \dot{\pvec{q}} &= \frac{\pvec p}{m},\\
    \dot{\pvec{p}} &= -\del_{\pvec q} V(\pvec q) - \frac{ \gamma}{m}\pvec p + \sqrt{2D_{\pvec p}}\,\pvec\xi.
\end{align}
    \label{eq:FullLangevin}
\end{subequations}
Eqs.~(\ref{eq:FullLangevin}) resemble the Hamiltonian equations of motion (\ref{eq:HamiltonianEOM}) with two additional (force) terms for the change of momenta.
The first additional term phenomenologically models the solvent friction with drag coefficient $\gamma$.
The second term is the stochastic force $\pvec \xi$ with strength characterized by $D_{\pvec p}$.
The statistics of the force $\pvec \xi$ are those of white noise,
\begin{equation}
  \eav{ \pvec \xi(t) } = 0, \qquad \eav{ \xi_i(t)\xi_j(t') } = \delta_{ij}\delta(t-t'),
  \label{eq:WhiteNoise}
\end{equation}
where $\xi_i$ denotes the $i$th component of $\pvec \xi$, $\delta_{ij}$ is the Kronecker-symbol and $\delta(t-t')$ is the Dirac $\delta$-distribution.
The $\delta$-distribution for the correlations is an approximation to the real collision statistics that requires that the time scale of observations $\tobs$ is much larger than the typical time of a microscopic collision, $\tmic$.
For the momenta to follow a Maxwell-Boltzmann distribution in the steady-state, $D_{\pvec p}$ has to obey the \emph{fluctuation-dissipation relation}
\begin{equation}
  D_{\pvec p} = \frac{\gamma}{m^2}\kb T,
  \label{eq:FDRMomenta}
\end{equation}
where $T$ is the temperature of the bath and $\kb$ is Boltzmann's constant.

The evolution of the probability density $\micden^{(t)}(x)$ of finding the system at time $t$ at phase space point $x \in \pspace_{\mathrm{sys}}$ is governed by the Fokker-Planck (or in this case, the so-called Kramers-Klein) equation
\begin{equation}
  \del_t\micden^{(t)}(x) = - \sum \left[ \del_{\pvec q} \pvec{j}_{\pvec q} + \del_{\pvec p}\pvec{j}_{\pvec p}\right]
  \label{eq:KramersKlein}
\end{equation}
with current densities
\begin{equation}
  \pvec{j}_{\pvec q} = \frac{\pvec p}{m}\micden^{(t)},\quad\pvec{j}_{\pvec p}=\left[ \left(\del_{\pvec q}V\right)-\gamma\frac{\pvec p}{m}+\gamma \kb T \del_{\pvec p} \right]\micden^{(t)}.
  \label{eq:CurrentDensitiesKK}
\end{equation}
Note that we use the notation introduced in \eq{HamiltonianEOM} in section \ref{sec:microscopic_basics}.

\subsection{Smoluchowski equation}
Often it is not feasible or necessary to explicitly consider the momenta $\pvec p$.
On average, the momenta reach the value $\eav{ \pvec p } = -\frac{m}{\gamma}\del_{\pvec q }V = \frac{m}{\gamma}\pvec F$ on a time scale $\tmom := \frac{m}{\gamma}$.
If the system's particles are atoms or small molecules, $\tmom \sim \tmic$.
If the system's particles are already mesoscopic objects and a (hydrodynamic) radius $R$ can be defined, Stokes friction yields $\tmom \propto \mu^{-1} R^{2}$ with the dynamic viscosity of the environment $\mu$.
For instance, for colloidal silica of a couple of hundreds of nanometers in diameter one has $\tmom \approx 10^{-11}s$ which is well below usual observable time scale $\tobs$.

In general, if $\tobs \gg \tmom$ the momenta are always relaxed to their equilibrium values and one can consider an \emph{overdamped dynamics} using the Smoluchowski equation generating dynamics on the space of configurations $\pspace_{\mathrm{conf}}$:
\begin{equation}
  \dot{\pvec {q}}= \mu \del_{\pvec q}V + \sqrt{2D}\pvec \xi.
  \label{eq:SmoluchowskiEquation}
\end{equation}
Here, $\mu = \frac 1 \gamma$ is the mobility and $D$ is the diffusion constant.
Using either the FDR (\ref{eq:FDRMomenta}) or demanding the steady-state distribution to be of Boltzmann form yields the so-called Einstein-Smoluchowski relation:
\begin{equation}
  D = \frac{\kb T }{\gamma m^2}=\frac{\mu}{m^2}\kb T.
  \label{eq:FDREinstein}
\end{equation}
To find a typical time-scale $\tcoord$ for the overdamped dynamics (\ref{eq:SmoluchowskiEquation}) suppose that the system is close to its equilibrium condition and therefore a harmonic approximation for the potential energy holds, \ie $V(\pvec q) \approx \frac k 2 \pvec q^2$ with some spring constant $k$ which is of the order of the typical interaction energy (for biological system a few $\kb T \approx 10^{-20} J$) divided by the square of the typical length-scale of the interaction (typically on the order of nanometers).
Then the force is linear in $\pvec q$ yielding a time scale $\tcoord = \frac{\gamma}{k}$.
Again, using Stokes friction one finds for typical values  $\frac{\tcoord}{\tmom} \sim \frac{\mu^2 R}{k} \gg 1$.
This justifies that we neglect the dynamics of the momenta in favour of the dynamics of the coordinates.

However, in typical single-molecule experiments on proteins or DNA $\tobs \gg \tcoord$.
In addition, spatial resolution is usually not good enough to resolve the dynamics of single coordinate degrees of freedom.
This is why one might be interested in much slower, collective dynamics happening on larger time- and length-scales which are accessible to experimental observation.
One example for this further separation of scales are configurational changes or folding in proteins occurring on time scales $\tconf \approx 10^{-4}$--$10^{0}s$.

\subsection{Stochastic dynamics of mesoscopic observables}
If the detail of observation on a system is restricted to certain meso- or macroscopic observables $\meas(x) = \meas(\pvec q)$, it is natural to ask for a (stochastic) description only involving the possible values $\omega\in\ospace$ that these observables can take.
As every configuration $\pvec q$ corresponds to exactly one observable state $\omega = \meas(x) \in \ospace$, the allowed values $\omega$ then partition $\pspace_{\mathrm{sys}}$ into disjoint classes (cells)
\begin{equation}
  \cell_{\omega} := \set{ {x \in \pspace_{\mathrm{sys}} \vert \meas(x) = \omega}}.
  \label{eq:Classes}
\end{equation}
We will distinguish two different situations.

\paragraph{Continuous observable}
Firstly, let $\meas$ be some continuous function of $x$.
The classes $\cell_{\omega}$ are then hypersurfaces in $\pspace_{\mathrm{sys}}$.
Often, in that case we can write down a phenomenological coupled Langevin equation for the stochastic variable $\omega$:
\begin{equation}
  \dot{{\omega}} = \uuline{\mu}\left[ -(\nabla F({\omega})) +  f \right] +  \zeta.
  \label{eq:LangevinMesoscopic}
\end{equation}
If $\omega$ is multi-dimensional, the mobility tensor $\uuline \mu$ couples the different components of $\omega$ and $\zeta$ is generalized white noise with correlation matrix $\eav{ \zeta(t) \zeta(t') } = 2 \kb T \uuline \mu\delta(t-t')$.
$F(\omega)$ is a phenomenological potential (which should be interpreted as a free energy, see below) and $f$ is a non-conservative force.

The non-conservative force can arise through the projection of $\pspace_{\mathrm{sys}}$ onto $\ospace$ if $\meas(x)$ is a non-monotonous function yielding classes $\cell_\omega$ which are not simply connected.
For instance, consider a description where $\meas(x) = \meas(\pvec q)$ is periodic in $\pvec q$, \ie where the effective dynamics is done using periodic boundary conditions to model an infinite system with an external force $F_{pot} = -\nabla V$.
In this case, the force governing the dynamics of $\omega$ turns into a non-conservative force $f$ acting on a torus.

The Fokker-Planck equation for \eq{LangevinMesoscopic} is
\begin{equation}
  \del_t \mesden^{(t)} = -\nabla j \equiv -\nabla\left(\uuline \mu \left[-(\nabla F({\omega})) +  f - \kb T \nabla  \right]\mesden^{(t)}\right).
  \label{eq:FokkerPlanckMesoscopic}
\end{equation}

\paragraph{Discrete observable}
Another possibility is that we already start with a disjoint discrete partition $\set{\cell_i}_{i=1}^M$ and assign values $\omega_i$ to each of its elements, \ie
\begin{equation}
  \meas(x) = \sum_{i=1}^{M} \left[\omega_i \chi_{\cell_i}(x)\right].
  \label{eq:DiscreteObservable}
\end{equation}
The classes  $\cell_{\omega_i} \equiv \cell_i$ are identical with the elements of the partition.

In this case the discrete dynamical trajectories $\traj\omega(t)$ are created by a \emph{Markov jump process} rather than by a Langevin equation.
The probabilities $p_i^{(t)}$ of being in state $i$ at time $t$ evolves according to the \emph{Master equation}, which is a discrete version of the Fokker-Planck equation:
\begin{equation}
  \del_t p_i^{(t)} = \sum_{j=1}^M \left[ w^j_i p_j^{(t)} - w^i_j p_i^{(t)} \right]
  \label{eq:MasterEquation}
\end{equation}
We assume that the jump rates $w^i_j\geq0$ for $i\neq j$  and $w^i_i = -\sum_j w^i_j$ are time independent.
As the original physical dynamics is time-reversible, each trajectory that leads the system from $\cell_i$ to $\cell_j$ must also have an allowed reverse trajectory.
This condition of \emph{dynamical reversibility} can only be fulfilled if jumps can always happen in both directions, \ie  $w^i_j>0 \Leftrightarrow w^j_i >0$.

One can also consider the jump dynamics as a random walk on a graph with $M$ nodes with edges $(i,j)$ if and only if $ w^i_j >0$.
This graph can have arbitrary topology including cycles, which are the analogue of the non-Euclidean (\eg toroidal) geometries for the continuous case.

\subsection{Local equilibrium and intrinsic entropy of observable mesoscopic states}
We just saw how a further separation of time-scales between $\tcoord$ and the typical time-scales of observable collective motion $\tconf\geq\tobs$ lead to effective mesoscopic descriptions of \eq{LangevinMesoscopic} and \eq{MasterEquation}.
Physically, to justify the Markovian stochastic dynamics on the level of the mesoscopic cells, one has to assume equilibrated cells.
This means that the dynamics happening within a cell reach a constrained equilibrium, which is, by definition, memoryless.
Local equilibrium is the key concept for the consistency of coarse-grained descriptions \cite{deGroot+Mazur1984,Falcioni_etal2007,Esposito2012}.
Following Seifert \cite{Seifert2011} we introduce the notions necessary in the present context:

Let $\meas$ describe some measurable observable that is to be treated within the framework of stochastic thermodynamics.
We assume that all (coordinate) microstates $\pvec q$ in any cell $\cell_\omega$ defined by this observable equilibrate on time-scales $\tcoord$ that are much shorter than the time-scales $\tconf$ of transitions between mesoscopic states.
With $\beta = (\kb T)^{-1}$, the conditioned microscopic distribution $\micden(\pvec q\vert\omega)$ of finding microstate $\pvec q$ given mesoscopic state $\omega$ reads
\begin{equation}
  \micden(\pvec q\vert\omega) = \exp\left( -\beta (V(\pvec q) - F(\omega) \right),
  \label{eq:ConstrainedProbability}
\end{equation}
where $V(\pvec q)$ is the microscopic potential on $\pspace_{\mathrm{conf}}$ and 
\begin{equation}
  \begin{split}
    F(\omega) :=& -\kb T \log \left( \sum_{\pvec q \in \cell_{\omega}}\left[ \exp\left( -\beta V\left(\pvec q\right) \right)\right] \right)\\ 
  \end{split}
  \label{eq:FreeEnergy}
\end{equation}
is the free energy of state $\omega$.
$F = e - T \intrv$ can be split into the constrained internal energy
 \begin{equation}
   e(\omega) = \langle V \rangle_{\omega}=\sum_{\pvec q \in \cell_{\omega}}\left[\micden(\pvec q \vert \omega) V(\pvec q)\right]    \label{eq:InternalEnergy}
 \end{equation}
 and the intrinsic entropy of state $\omega$,
\begin{equation}
  \intrv(\omega) =-\kb \sum_{\pvec q \in \cell_{\omega}}\left[ \micden(\pvec q\vert\omega) \log \micden(\pvec q\vert\omega)\right].
  \label{eq:IntrinsicEntropyStochastic}
\end{equation}

Here, we formulated everything only with coordinates $\pvec q$.
In underdamped cases, usually a Maxwell-distribution for the momenta $\pvec p$ is assumed, which corresponds to a fast equilibration of the momenta.
The difference that would occur both in the intrinsic entropy and free energy is then only an (unobservable) constant.
For more mathematically rigorous statements on the general issue of considering entropies of reduced descriptions see Ref.~\cite{Maes+Netocny2003}.

\subsection{Mesoscopic trajectories and trajectory averages}
The goal of stochastic thermodynamics is to formulate sensible thermodynamic balance equations on the level of stochastic trajectories $\traj \omega$.
In this context we will consider mesoscopic trajectories that are discrete in the space of observations but continuous in time.
A similar treatment for the case where $\omega$ is a continuous variable can be found in Refs.\,\cite{Seifert2005,Seifert2011,Seifert2012}.

A trajectory $\traj \omega$ is a generated by a Markov jump process, \ie it is a random variable.
Almost all trajectories can be described by two countably infinite sequence:
\begin{equation}
  \traj \omega = \left(
  \begin{array}[c]{ccc}
  \omega_0 &\omega_1&\ldots\\
  \tau_1&\tau_2&\ldots
  \end{array} \right).
  \label{eq:InfiniteContinuousTrajectory}
\end{equation}
In the first row the different states visited by the trajectory are listed.
The second row contains the jump times $\tau_j$ of the $j$th jump occurring between states $\omega_{j-1}$ and $\omega_j$.
A finite trajectory $\traj\omega{(\tau)}$ runs only for times $t \in [0,\tau]$.
The number of jumps $n = n(\tau)$ is a random variable.
The finite trajectory $\traj \omega{\left( \tau \right)}$ can be described by two vectors of length $n+1$:
\begin{equation}
  \traj \omega = \left(
  \begin{array}[c]{cccccc}
    \omega_0 &\omega_1&\ldots&\omega_{n(\tau)-1}&\omega_{n(\tau)}\\
    \tau_1&\tau_2&\ldots&\tau_{n(\tau)}&\tau
  \end{array} \right).
  \label{eq:FiniteContinuousTrajectory}
\end{equation}
The last entry in the second row is the temporal length of the trajectory, $\tau$.
In both cases, we denote the mesoscopic state at time $t \in [0,\tau]$ along a trajectory by $\omega(t)$.

The weight of a trajectory of length $\tau$ on the appropriate trajectory space is denoted by $\mathbb P[\traj\omega]$.
Henceforth, to stress a functional dependence we will use square brackets $[\cdot]$.
For instance for the formulation of fluctuation theorems, it is useful to split the trajectory weight into an initial and a conditional part:
\begin{equation}
  \mathbb P\left[ \traj \omega \right] = p^{\left( 0 \right)}_{\omega_0}\cdot p\left[\traj \omega \mvert \omega_0\right]
  \label{eq:TrajectoryWeightSplitting}.
\end{equation}

Let $\obs[\traj\omega;\tau]\equiv \obs\left[ \omega(\tau);\tau \right]$ be a functional of trajectories running for time $\tau$ which can also explicitly depend on $\tau$.
The time-dependent trajectory average is defined as
\begin{equation}
  \trav{\obs} := \sum_{\traj \omega (\tau)}\left[\mathbb P[\traj \omega(\tau)] \obs[\traj\omega(\tau);\tau]\right].
  \label{eq:TrajectoryAverageGeneral}
\end{equation}
If $\obs \left[ \traj \omega; \tau \right] = \obs(\omega(\tau);\tau)$ only depends on the final state $\omega(\tau)$ of the trajectory, we say it has a \emph{local} form.
Local forms obey
\begin{equation}
  \trav{\obs} = \sum_{i}\left[p_i^{\left( \tau \right)} \obs(\omega(\tau);\tau)\right] = \eav{\obs}^{(\tau)},
  \label{eq:AverageOfLocalObservable}
\end{equation}
\ie they can be expressed as trajectory averages.
A special case of local forms are state-variables, where $\obs(\omega)$ denotes a state function in the sense of classical thermodynamics.
Examples of non-local physical observables are current variables $\obs[\traj\omega;\tau] = \sum_k\left[ \obs (\omega_{k-1},\omega_k;\tau) \delta (\tau-\tau_k)\right]$ which only depend on single jumps $\omega_{k-1} \to \omega_k$ that occur at time $\tau_k$.
For such observables, the trajectory averages reduce to current averages which involve summing over two states (hence the subscript $_2$):
\begin{equation}
  \trav{\obs} = \sum_{i,j}\left[p_i^{\left( \tau \right)}w^i_j \obs(i,j;\tau)\right] = \eav{\obs}^{(\tau)}_2.
  \label{eq:AverageOfCurrentObservable}
\end{equation}

\subsection{Thermodynamic balance}
\paragraph{Single jumps}
As we will see below, the entropic trajectory functionals of stochastic thermodynamics are all linear combinations of three fundamental functionals, which are either state variables, local forms or current variables.
Here, implicit time-dependence enters through an explicit dependence on the solution of the equation for the evolution of the ensemble probabilities.
The same ideas also apply if the transition rates explicitly depend on time.
%
We start by considering the changes associated with a \emph{single jump} from $\omega_i$ to $\omega_j$ happening in an infinitesimal time interval.
We will denote the changes encountered at a jump by the symbol $\df$ for state variables, $\Delta$ for local observables and $\delta$ for current variables.

The change in internal energy is given by $\df e = e(\omega_j) - e(\omega_i)$.
If additionally a non-conservative (generalized) force $f$ is acting an additional amount of (mechanical) work $\delta w$ is dissipated.
The change of heat $\delta q = \df e + \delta w$ in the medium consists of the (negative) change of internal energy plus the dissipated energy.
Therefore, the change of entropy in the medium is given as
\begin{equation}
  \delta \medrv = \frac{\delta q}{T} = \frac{\df e}{T} + \frac{\delta w}{T}.
  \label{eq:DefnMediumEntropy}
\end{equation}

To formulate the second law we need to identify all terms corresponding to the \emph{change of the total entropy}, $\delta \totrv$, in system and environment.
However, it is not enough to identify the sum of $\delta \medrv$ and $\Delta \intrv = \intrv(\omega_j) - \intrv(\omega_i)$ with $\delta \totrv$. 
If it were so, one can easily construct a violation of the second law where $\delta \totrv$ is not positive on average \cite{Seifert2011}.%
\footnote{The argument of Ref. \cite{Seifert2011} goes as follows: Consider a system without external forces and two states with the same internal energy but different intrinsic entropies.
An ensemble that initially consists only of systems in the state with the higher intrinsic entropy would be smeared out over both states while decreasing its entropy in the process.}

Therefore another term is needed to account for the current state of the ensemble.
This term will be of local form but not a state variable, hence we denote it by $\Delta \visrv$.
The subscript indicates that this is the \emph{visible} entropy, which an experimenter could measure by sampling the ensemble.

Putting everything together, we find for a jump from $\omega_i$ to $\omega_j$
\begin{equation}
  \delta \totrv(i\to j) = \delta \medrv(i\to j) + \df \intrv(i\to j) + \Delta \visrv(i\to j).
  \label{eq:TotalEntropyBalance}
\end{equation}

The instantaneous average rate of change of the total entropy, $\totep$, is obtained by summing over all jumps with the corresponding joint probability rates and yields a dynamical formulation of the second law:
\begin{equation}
  \totep(t) := \sum_{i,j}\left[ p_i^{(t)} w^i_j \delta \totrv(i\to j)\right]\stackrel{!}{\geq} 0
  \label{eq:SecondLawDynamical}
\end{equation}
$p_i^{(t)}$ is the solution of the Master equation (\ref{eq:MasterEquation}) at time $t$ for some given initial condition.

The only thermodynamically consistent choice for $\totep$ that fulfils the second law is a form resembling a Kullback-Leibler divergence \cite{Seifert2012}.
It has frequently been used in previous literature (\cf Refs.~\cite{Schnakenberg1976,Gaspard2004}):
\begin{equation}
  \totep(t) := \kb \sum_{i,j}\left[ p_i^{(t)} w^i_j \log \frac{p_i^{(t)}w^i_j}{p_j^{(t)}w^j_i} \right] \geq 0.
  \label{eq:TotalEP}
\end{equation}
The stochastic entropy can now be identified with the ensemble-dependent part in the above equation:
\begin{equation}
  \Delta \visrv := -\kb \log p_j^{(t)} - \left(-\kb \log p_i^{(t)} \right) \label{eq:stochasticEP}.
\end{equation}
The remaining part summarizes the other two contributions:
\begin{equation}
  \delta\jumprv = \delta \medrv + \df \intrv = \kb \log \frac{ w^i_j}{w^j_i} =: B^i_j.
\end{equation}
To simplify notation, we introduce the quantity
\begin{equation}
  B^i_j := \kb \log \frac{ w^i_j}{w^j_i}    \label{eq:Electromotance}  
\end{equation}
also previously defined in the early works of Hill \cite{Hill1977}, in the context of fluctuation relations for stochastic dynamics \cite{Kurchan1998,Lebowitz+Spohn1999,Seifert2005,Andrieux+Gaspard2007,Faggionato+dPietro2011} and related works \cite{Altaner_etal2012, Altaner+Vollmer2012}.
We suggest to call $B^i_j$ (local) \emph{motance}, hence also the name of the corresponding entropy, $\jumprv$.
It is the ability of the physical system (independent of the ensemble) to facilitate thermodynamic motion.
It usually corresponds to changes in the natural thermodynamic potentials of certain physical situations \cite{Seifert2011}.

\paragraph{Balance along trajectories}
We want to extend the balance for a single jump to a balance involving trajectory-dependent functionals, which are an easy approach to the fluctuation relations for stochastic dynamics.
Instead of starting with the physical functionals for the entropy production in system and medium, we identify three fundamental functionals with distinct properties, which constitute the building blocks of all other functionals.
The first functional is related to the intrinsic entropy, which is a state variable.
Consequently, this functional is independent of the ensemble and does only depend on the end-point $\omega(\tau)$ of a trajectory $\traj \omega (\tau)$:
\begin{equation}
  \intrv[\traj \omega;\tau] = \intrv\left(\omega\left(\tau\right)\right) = -\kb \sum_{\pvec q \in \cell_{\omega(\tau)}} \left[\micden(\pvec q\vert\omega(\tau)) \log \micden(\pvec q\vert\omega(\tau))\right].
  \label{eq:IntrinsicEntropyTraj}
\end{equation}
Unfortunately, one never has direct access to $\intrv$ because the intrinsic entropy is not visible in the coarse-grained description.
However, we can measure the ensemble-dependent visible entropy, which is also a local functional:
\begin{equation}
  \visrv[\traj \omega;\tau] = \visrv(\omega(\tau))= - \kb \log p_{\omega(\tau)}^{(\tau)},\label{eq:VisibleEntropyTraj}  
\end{equation}

The entropy associated with the motance $\delta \jumprv$ cannot be written in difference form. 
It depends on the jumps along the trajectory, so the corresponding functional is a current variable an will depend on the whole trajectory $\traj \omega(\tau)$.
We introduce the motance of a trajectory $\jumprv$, which needs to integrate all jumps from state $\omega_{k-1}$ to state $\omega_k$ that have occurred at times $\tau_k<t$:
\begin{equation}
  \jumprv[\traj \omega;\tau] = \kb \sum_k\left[ \Theta(\tau-\tau_k) B^{\omega_{k-1}}_{\omega_k}\right].
  \label{eq:JumpEntropy}
\end{equation}
Mathematically, the motance is the time-antisymmetric part of the action Lagrangian in the sense of an action functional for trajectories \cite{Maes2004,Seifert2012}.
It consists of a part corresponding to a heat released to the medium and a change in intrinsic entropy.
From those quantities one can construct the functionals that quantify the entropy change in the system and medium, $\sysrv[\traj\omega]$ and $\medrv[\traj \omega]$,
\begin{align}
  \sysrv[\traj\omega]&= \visrv[\traj\omega] + \intrv[\traj\omega]\label{eq:SystemEntropyTraj},\\
  \medrv[\traj\omega] &= \jumprv[\traj\omega] -\intrv[\traj\omega],\\
  \totrv[\traj\omega] &= \medrv[\traj\omega] + \sysrv[\traj\omega] = \jumprv[\traj\omega] + \visrv[\traj\omega]\label{eq:TotalEntropyTraj}.
\end{align}

The instantaneous rates of change of the entropies have previously been identified \cite{Seifert2005}.%
\footnote{Here, to avoid dealing with time derivatives of random variables, we do not denote them by $\dot s$ but rather by the symbol $\sigma$.}
\begin{align}
  \viseprv[\traj \omega] &= - \frac{\del_\tau p_\omega(t)\vert_{t= \tau, \omega =\omega(\tau)}}{p_{\omega{(\tau)}}^{(\tau)}} - \sum_k\left[ \log\left( \frac{p_{\omega_k}^{(\tau)}}{p_{\omega_{k-1}}^{(\tau)}}\right)\delta(\tau-\tau_k)\right],\label{eq:VisEntropyRateTraj} \\
  \inteprv[\traj\omega] &=  \sum_k\left[\left(\intrv(\omega_k) - \intrv(\omega_{k-1})\right)\delta(\tau - \tau_k)\right],\\
  \jumpeprv[\traj\omega] &= \kb \sum_k \left[B^{\omega_{k-1}}_{\omega_k} \delta(\tau-\tau_k)\right] \label{eq:JumpEntropyRateTraj}. 
\end{align}

\subsection{Averaging over trajectories}
Strictly speaking, of the above defined entropic quantities only the intrinsic entropy is a state variable and therefore has a counter-part in classical thermodynamics.
However, upon averaging over trajectories, we recover the thermodynamic interpretations of Markov processes originally used by Hill and later Schnakenberg \cite{Hill1977,Schnakenberg1976}.
We start with the visible entropy.
As a local quantity, it reduces to an ensemble average which is the usual Shannon entropy of the (discrete) ensemble:
\begin{equation}
  \visent(\tau) := \trav{ s_{vis}} = \eav{ s_{vis}}_\tau =  -\kb\sum_i\left[ p_i^{(\tau)} \log p_i^{(\tau)}\right]
  \label{eq:ShannonEntOfEnsemble}
\end{equation}
Let us look at the time derivative of this entropy:
\begin{equation}
  \visep(\tau) :=\tdiff{\visent}{\tau}\stackrel{(\ref{eq:MasterEquation})}{=}\kb \sum_{i,j}\left[ p^{\left( \tau \right)}_i w^i_j \log\left. \frac{p_i^{\left( \tau \right)}}{p_j^{\left( \tau \right)}} \right.\right] = \trav{\viseprv}.
\end{equation}
Because the average is linear, we can also write it as the difference of the trajectory averages of the terms $\toteprv$ and $\jumpeprv$, which can also be explicitly calculated.
\begin{align}
  \visep(\tau) &= \trav{ \toteprv} - \trav{\medeprv} \nonumber\\
  &= \kb \sum_{i,j} \left[p_i^{(\tau)} w^i_j \log \frac{p_i^{(\tau)}w^i_j}{p_j^{(\tau)}w^j_i}\right] -\kb\sum_{i,j}\left[ p_i^{(\tau)} w^i_j \log \frac{w^i_j}{w^j_i}\right]\\
                  &=: \totep(\tau) - \jumpep(\tau).
  \label{eq:functionalsToSchnakenberg}
\end{align}
Often, $\visep$ and $\jumpep$ are identified with the entropy production rate in the system and environment, respectively, which is only true if we agree to forget about the change in intrinsic entropy.
However, in a steady state the average intrinsic entropy is constant and such an identification is correct again (though trivial, because $\visep=0$).

\subsection{Summary}
In this section we have reviewed the basic concepts of stochastic thermodynamics.
The main result was the identification of the three fundamental entropy functionals \intrv, \visrv and \jumprv together with the functionals for the variations, \inteprv, \viseprv and $\jumpeprv$.
The crucial, yet somewhat uncommon observation is that the trajectory functionals can depend on the current ensemble, \ie the solution of the ensemble evolution equation.
Further, the functionals were shown to be consistent with older results on the thermodynamic interpretation of Markov processes.
We deliberately skipped the application of the approach, like the derivation of fluctuation theorems, which can be found in the literature \cite{Maes2004,Seifert2005,Seifert2012}.

\section{Deterministic Thermostatted Dynamics}
\label{sec:thermostats}
In the last section we have outlined the steps of how to arrive from a Hamiltonian dynamics at trajectory-dependent functionals for entropic terms of stochastic thermodynamics.
Stochasticity  was introduced from the beginning to deal with the ignorance of certain aspects of the system which lie beyond the scope or resolution of our observations.
The rest of this paper is dedicated to accomplishing the same using deterministic dynamics.

In this section we will outline the same conceptual steps as in the previous section only in the deterministic framework:
We will review thermodynamic consistency, reduction to overdamped dynamics and finally the coarse-graining to a mesoscopic description.

\subsection{General scheme}
Throughout this section, we will follow the systematic approach to deterministic thermostats presented in Ref. \cite{Samoletov_etal2007}.
This general scheme resembles the formulation of a Langevin equation:
One starts with an equation of motion for the degrees of freedom of the system under consideration.
After that, a phenomenological drag term is added to model the influence of the environment.
Instead of adding a stochastic noise term, one either promotes the drag coefficient to a dynamical variable or adds a deterministic ``noise'' term mimicking the fluctuations caused by the environment.
As in the stochastic case, the choice of these terms has to be consistent with the thermodynamic properties of the bath.
This is achieved by demanding that the stationary distribution $\micden^{\infty}$ of the physical degrees of freedom obtains a canonical form.

\subsection{Nos\'e--Hoover thermostats}
We will exemplify the above method using the Nos\'e--Hoover scheme, which is the deterministic analogue to the full Langevin equation \eq{FullLangevin}.
The equations of motion are
  \begin{subequations}
\begin{align}
    \dot{\pvec{q}} &= \frac{\pvec p}{m},\\
    \dot{\pvec{p}} &= -\del_{\pvec q} V(\pvec q) - \frac{\widetilde \gamma}{m}\pvec p,\\
    \dot{\widetilde \gamma}  &= g(\pvec q,\pvec p).
\end{align}
    \label{eq:NoseHoover}
  \end{subequations}

Demanding that
\begin{equation}
  \micden^{\infty}\propto \exp\left( -\beta\left(V(\pvec q)+ \sum\left[\frac{\pvec p ^2}{2m}\right] +\Phi(\widetilde \gamma) \right) \right)
  \label{eq:NoseHooverStationary}
\end{equation}
with a quadratic term $\Phi(\widetilde \gamma)$ (\ie Gaussian distribution of the values of $\widetilde \gamma$) one arrives at
\begin{align}
  \Phi(\widetilde \gamma) &= \frac{Q}{2m^2}\dot{\widetilde \gamma}^2 \label{eq:NoseHooverPhi},\\
  g(\pvec q,\pvec p) &= \frac{m}{Q}\left( \sum \left[\frac{\pvec p^2}{m}\right]-d N \kb T \right) .
  \label{eq:NoseHooverG}
\end{align}
Here, $d$ is the dimension of physical space and $Q=d N \kb T \tmom^2$ is a constant related to the time scale $\tmom$ of the relaxation of the momenta.
The discussion in the previous section motivates $Q = d N \kb T \frac{m^2}{\gamma^2}$, where $\gamma$ is the phenomenological drag constant.
If such a drag constant is not known, $Q$ is a free parameter of the dynamics.

Time averaging of equation \eq{NoseHooverG} in the sense of \eq{TimeAverageInfinity} leads to the relation
\begin{equation}
  \bar{\left( \sum \frac{\pvec p^2}{m} \right)}=d N \kb T.
  \label{eq:Equipartition}
\end{equation}
It describes equipartition of the momenta coordinates at temperature $T$.
Therefore, the dynamics of $\widetilde \gamma$ ensures that the momenta relax to their equilibrium values.%
\footnote{As in the stochastic case, it is an assumption that the trajectories generated by the thermostatted equations of motion can describe the transient relaxation towards equilibrium, too.}

\subsection{Configurational thermostats}
Configurational thermostat are the deterministic analogue to the overdamped Smoluchowski equation (\ref{eq:SmoluchowskiEquation}).
It is assumed that the momenta have relaxed to their equilibrium values and the dynamics can be described by the equations
\begin{subequations} 
  \begin{align}
    \dot{\pvec {q}} &= \widetilde \mu \,\del_{\pvec q}V,
    \label{eq:OverdampedThermostat}\\
    \dot{\widetilde \mu} &= \frac{1}{Q_{\mu}}\sum\left[ \left( \del_{\pvec q}V \right)^2-\kb T \del_{\pvec q}^2 V \right].
    \label{eq:OverdampedG}
  \end{align}
  \label{eq:Overdamped}
\end{subequations}
Again, the form of the dynamics of $\widetilde \mu$ is found from demanding the canonical form for the stationary distribution of the coordinates.
Observe also how here we find the definition of Rugh's configurational temperature \cite{Rugh1997} appearing naturally in \eq{OverdampedG}.
As above, the constant $Q_{\mu}$ can be used to set a time-scale for relaxation of coordinates.

The problem with this approach is that the dynamics is not ergodic.
Mechanical equilibria $\del_{\pvec q}V = 0$ act as attracting fixed points where the system comes to rest.
To restore ergodicity, further modifications of the equations of motion are required.
In the so-called SDC scheme this is done by introducing another dynamical variable, $\pvec \zeta$, that is used to shake the dynamics around mechanical equilibria, similar to what the stochastic noise term would do in the Langevin equation \cite{Samoletov_etal2007}.
The equations of motion are
\begin{equation}
    \dot{\pvec {q}} = \mu \del_{\pvec q}V + \pvec \zeta
  \label{eq:SDCthermostat}
\end{equation}
with the dynamics of $\mu$ as in \eq{OverdampedG} and
\begin{equation}
  \dot{\pvec \zeta} = h(\pvec \zeta, \pvec q)
\end{equation}
with $h$ leading to a dynamics that satisfies $\pvec \zeta \cdot \del_{\pvec q} V = 0$.
There are essentially three different possibilities to satisfy this condition which correspond to differently constrained fluctuations around mechanical equilibria.

Actually, the full SDC scheme goes further and introduces a dynamical drag term for the coordinates, too.
If this is done, the connection to statistical physics here is obtained similarly to the Nos\'e-Hoover case, only that one uses equipartition for the virial $\sum\pvec q \cdot \del_{\pvec q}V(\pvec q)$ rather than for the mean kinetic energy.

It has not been proven if or under which conditions the dynamics created by the SDC thermostatting scheme are ergodic.
However, numerical simulations using deterministic SDC schemes as well as a variant \cite{Samoletov_etal2010}, the Braga-Travis (BT) scheme \cite{Braga+Travis2005}, indicate ergodicity.

\subsection{Dynamic reversibility}
A very important aspect of the SDC dynamics is their dynamic reversibility.
Dynamic reversibility means that we can basically let the dynamics run backward in time if we apply a time-reversal operator $\mathcal I: \pspace \to \pspace$ to its microstates.
For later reference we now formulate the general definition of reversibility for a dynamical system on phase space $\pspace$ governed by the flow $\Phi^{t}$, \eq{DefinitionFlow} \cite{Maes+Netocny2003,Maes2004,Jepps+Rondoni2010}.
Here, $\pspace$ is the extended configurational phase space with elements $x = (\pvec q, \alpha)$ consisting of the coordinates as well as the additional dynamical variables $ \alpha = (\mu, \pvec \zeta, \ldots )$.
Now let $\left( \pspace,\mathcal B, \lambda\right)$ be the measure space of the thermostatted system with Lebesgue measure $\lambda$ on the Borel sets $\mathcal B$.

We call the dynamics \emph{reversible} if and only if there is a mapping $\mathcal I$ such that
\begin{eqnarray}
  \mathcal I\circ\mathcal I = \text{id} & &\text{(involution)}, \label{eq:defnInvolution}\\
  \lambda(\mathcal I^{-1}A) = \lambda(A),~\forall A\in \mathcal B & &\text{(measure-preserving)} \label{eq:defnMeasurePreserving},\\
  \Phi^{(-t)} x = \left(\mathcal I \circ \Phi^{(t)}\circ \mathcal I\right) x,~\forall t\in\mathbb R, x \in \pspace. & & \text{(time reversal)}
  \label{eq:defnReversibility}
\end{eqnarray}

In the case of the SDC scheme the time-reversal operator that fulfills the above properties is $\mathcal I (\pvec q, \alpha) = (\pvec q,-\alpha)$.
For Hamiltonian dynamics, it is $\mathcal I (\pvec q, \pvec p) = (\pvec q,-\pvec p)$.
But there are also other kind of abstract dynamics, like multi-baker maps, where such a condition of reversibility holds (for a review \cf Ref. \cite{Vollmer2002}).

\section{Dynamical Notions of Entropy and Information}
\label{sec:information_entropy}
In this section we will use reversible deterministic (thermostatted) dynamics to derive analogues of the trajectory-dependent entropic expressions (\ref{eq:VisibleEntropyTraj}-\ref{eq:JumpEntropyRateTraj}) using the concepts of information theory.
Unlike the previous sections, which mostly reviewed existing results, the work in this section is original though the basic notions were inspired by works of Vollmer and co-workers and others \cite{Breymann_etal1998,Vollmer_etal1998,Vollmer2002,Maes+Netocny2003}

The outline is as follows:
First we will partition the phase space of the system into disjoint mesoscopic cells $\cell_\omega$.
As in the stochastic case of section \ref{sec:stochastic_thermodynamics}, these cells correspond to the possible observations (measurements) $\omega \in \ospace$.
Our information of the system on the mesoscopic level is quantified by an evolving coarse-grained density $\mesden$.
In parallel, the evolution of a microscopic density $\micden$ is considered, which contains the full information about both state and history of the system.
The (experimental) uncertainty of the state of the system can be expressed as the relative information or Kullback-Leibler divergence of the two densities.
In order to understand the interpretation and evolution of the information-like quantities in the context of statistical physics, we need the evolution of the densities.
Therefore, an analogue to the crucial assumption of {\it equilibrated cells} or {\it local equilibrium} needs to be formulated.
Additionally, the fact that certain fast or auxiliary variables cannot be observed imposes a time-reversal-symmetry condition on the mesoscopic states.

\subsection{Mathematical set-up}
Starting point for our discussion is a reversible, deterministic dynamics on some space $\pspace$.
As a paradigm we can think of the flow $\Phi^{(t)}$ generated by the thermostatted equations of motion (\ref{eq:SDCthermostat}).

In order to account for the finite time resolution of real experiments, we take a {\it stroboscopic} point of view on the dynamics.
Mathematically, this means that from the continuous dynamics we obtain a deterministic, discrete map $\Phi \equiv \Phi^{\tau}$
\begin{equation}
  \begin{split}
    \Phi \colon \,\, \pspace &\to \pspace,\\
  x^{(\nu)} & \mapsto x^{(\nu+1)}
  \end{split}
  \label{eq:evolutionOperator}
\end{equation}
by fixing a small time-step $\tau=\tobs$ and only look at the microscopic state of the system at intervals of $\tau$.
After that, we could let $\tobs$ tend to zero to arrive at a continuous description.
In terms of Markov processes we obtain a Markov jump process described by a Master equation (\ref{eq:MasterEquation}) from a discrete time Markov chain.
However, the physical assumption of separation of time-scales still applies, so one must not assume validity of the continuous description for too small time scales $\tau\sim\tmic$.
Henceforth, the index for the discrete time is denoted $\nu\in \mathbb Z$ corresponding to $t=\tau\nu$.

\subsubsection{Observables and partitioning}

In order to connect to the statistical mechanics of mesoscopic systems we partition phase space into discrete cells $\cell_i \subset \pspace$  through observables of the form of \eq{DiscreteObservable}. 
A crucial condition on the measurable observables \meas is that they are invariant under time-reversal, \ie a single measurement cannot distinguish the direction of time:
\begin{equation}
  \meas(\mathcal I x) = \meas( x).
  \label{eq:ObservableReversibility}
\end{equation}
Physically, this excludes unobservable phenomena from the definition of our mesoscopic states.
Thinking of the SDC-thermostats this is the condition that the observables should only depend on the physical coordinates $\pvec q$ and not on the auxiliary variables $\alpha$.
In the underdamped situation this means that observables shouldn't depend on the fast momenta.
However, this condition also applies for the mesoscopic cells that constitute the multi-baker maps used by Vollmer \cite{Breymann_etal1998,Vollmer2002}.

To continue, we will need the phase space volume $\vcell_i$ of phase space cell $\cell_i$, which will be its Lebesgue measure.
If the (extended) phase space $\pspace$ has a finite measure, so will the cells.
The real space with coordinates $\pvec q$ is always considered to be confined to a finite volume, hence it has a finite measure.
However, this does not need to be the case for the momenta or auxiliary variables.
But because these variables are unobservable, cells will always be direct products of a finite observable configurational part, $\cell_{\mathrm{obs},\omega}$ and possible infinite parts $\cell_{\mathrm{unobs}}$ which do not depend on the value of $\omega$.

Therefore, when we write $\vcell_i = \lambda(\cell_i)$ we mean the measure projected onto the observable, configurational phase space.
Formally, let $\left( \pspace_{\mathrm{conf}},\mathcal B, \lambda\right)$ be the measure space of the (extended) thermostatted system with Lebesgue measure projected on the Borel sets $\mathcal B$ of the finite, observable part $\pspace_{\mathrm{conf}}$.

Having ensured finiteness of the measures, we can establish some further definitions:

Let $\cell^m_n \subset \cell_m$ be the set of all points that are mapped from cell $m$ to cell $n$.
Denote $s^m_n$ its relative volume in cell $m$.
Similarly, let $\tilde {\cell}^m_n \subset \cell_n$ be the set of all points in cell $n$ with pre-images in cell $m$ and $\tilde s^m_n$ its relative volume:
%
\begin{subequations}
  \begin{align}
    \cell^m_n &:= \{ x \in \cell_m: \Phi x \in \cell_n\}\\
    \vcell^m_n &:= \lambda(\cell^m_n )\\
    s^m_n &:= \frac{\vcell^m_n}{\vcell_m}
    \label{eq:defnS}
  \end{align}
  \begin{align}
    \tilde \cell^m_n &:=\{  y \in \cell_n: \Phi^{-1}  y \in \cell_m\}\\
    \tilde \vcell^m_n &:=  \lambda(\cell^m_n)\\
    \tilde s^m_n &:= \frac{\tilde \vcell^m_n}{\vcell_n}
    \label{eq:defnTildeS}
  \end{align}
  \label{eq:defnPhaseSpaceFractions}
\end{subequations}
Because the cells form a disjoint partition one has
\begin{equation}
  \sum_n s^m_n = \sum_n \tilde s^n_m = 1,\, \forall m.
  \label{eq:PhaseSpaceConservation}
\end{equation}

\subsubsection{Images and pre-images under reversible evolution}
We assume that the discrete dynamics is reversible with an Involution $\mathcal I$ in the sense of equations (\ref{eq:defnInvolution}--\ref{eq:defnReversibility}).
Because of the constraint \eq{ObservableReversibility} the involution is necessarily local on cells, meaning that mesoscopic cells are invariant sets under the involution:
\begin{equation}
  \mathcal I \cell_n = \cell_n,\,\forall n.
  \label{eq:defnLocalOnCells}
\end{equation}
With this relation and the notion of dynamical reversibility (\ref{eq:defnInvolution}--\ref{eq:defnReversibility}) it follows that
\begin{equation}
  \tilde s^m_n = s^n_m.
  \label{eq:volumeFractionsReversibility}
\end{equation}
This reason for this is that 
\begin{equation}
  \tilde s^m_n = s^n_m \Leftrightarrow \tilde \vcell^m_n  \equiv  \lambda(\tilde \cell^m_n) =  \lambda(\cell^n_m) \equiv \vcell^n_m 
\end{equation}  
by definition and further one has
\begin{align*}
  \tilde \cell^m_n &= \,\,\set{ y: (y \in \cell_n)\wedge(\phi^{-1}y \in \cell_m)}\\
  \Rightarrow \mathcal I \tilde \cell^m_n &=\, \,\set{ z: (z = \mathcal I y)\wedge(y \in \cell_n)\wedge(\phi^{-1}y \in \cell_m)}\\
  &\stackrel{\hidewidth(\ref{eq:defnReversibility})\hidewidth}{=}\,\, \set{ z: (\mathcal Iz \in \cell_n)\wedge(\phi^{-1}\mathcal I z \in \cell_m)}\\
  &\stackrel{\hidewidth(\ref{eq:defnInvolution})\hidewidth}{=} \,\,\set{ z: (z \in \mathcal I \cell_n)\wedge(\mathcal I\phi^{-1}\mathcal I z \in\mathcal I \cell_m) }\\
  &\stackrel{\hidewidth(\ref{eq:defnReversibility},\ref{eq:defnLocalOnCells})\hidewidth}{=} \,\,\set{ z: (z \in \cell_n)\wedge(\phi z \in \cell_m)}\\
  &\equiv \,\,\cell^n_m,
\end{align*}
which means that the role of images and pre-images is exchanged under the reversed dynamics.
Because the involution is measure preserving (\ref{eq:defnMeasurePreserving}) we find
\begin{equation}
  \vcell^n_m = \lambda(\cell^n_m) = \lambda(\mathcal I \tilde \cell^m_n) = \lambda(\tilde \cell^m_n) = \tilde \vcell^m_n
\end{equation}
which proofs \eq{volumeFractionsReversibility}.

\subsection{Information theory and entropy}
To connect the deterministic dynamics on the cells to thermodynamics we define entropic quantities motivated in the context of Shannon's {\it information theory} \cite{Shannon1948}.
We equip the space $(\pspace,\mathcal B)$ with a family of measures $\mu^{\left( \nu \right)}$ that are parametrized by discrete time $\nu$.
We assume that any member of the family $\mu^{\left( \nu \right)}$ has a density $\micden^{\left( \nu \right)}( x)$ with respect to the Lebesgue measure $\lambda$.
We will denote $\micden^{(\nu)}$ the  {\it fine-grained (or microscopic)} density.

The central quantity of information theory is the information functional of a density $\micden$ over an area of phase space $\cell \subset \pspace$:
\begin{equation}
  \sent{\micden\mvert\cell}:=-\kb \int_\cell \micden(x)\log\micden(x) \df x
  \label{eq:ShannonEntropy}
\end{equation}
The standard Shannon entropy is the information functional evaluated over the total phase space, $\sent \micden = \sent {\micden\mvert\pspace}$.
Another quantity that will appear frequently is the relative information of density $\micden'$ with respect to density $\micden$, which can be formulated as a Kullback-Leibler divergence:
\begin{equation}
  \kldiv{\micden\mVert \micden'}=\kb \int_\cell \micden(x)\log\left. \frac{\micden(x)}{\micden'(x)} \right. \df x > 0.
  \label{eq:Kullback-Leibler-Divergence}
\end{equation}

\subsubsection{Coarse-grained density}
Additionally to the microscopic density, we define a {\it coarse-grained (or mesoscopic)} density $\mesden^{(\nu)}_m$ to be the averaged microscopic density of a cell $\cell_m$:
\begin{equation}
  \mesden^{\left( \nu \right)}_m = \frac 1 {\vcell_m} \int_{\cell_m}\micden^{\left( \nu \right)}( x) \mathrm{d} x.
  \label{eq:coarseGrainedDensity}
\end{equation}
It resembles the fact that we can only measure up to a certain resolution that is defined by the mesoscopic observables $\meas(x)$.
That is, we can see a system in a mesostate $\omega$ but never infer the actual microstate $x$ from that information.

\subsubsection{Fine- and coarse-grained entropy}
For simplicity, in this paragraph we will suppress the superscript $\,^{\left( \nu \right)}$ because we are not dealing with any evolution of the densities.
We will refer to the constrained Shannon entropy of $\micden$ on cell $\cell_m$,
\begin{align}
    \sent{\micden \mvert \cell_m} &= -\kb\int_{\cell_m}\micden( x)\log\left.\micden(x)\right.\df x\label{eq:cellFGEntropy}\\
          &= -\kb\int_{\cell_m}\micden( x)\log \left( \vcell_m \micden( x) \right)\df x 
             +\kb\int_{\cell_m}\micden( x)\log \left. \vcell_m \right.\df x\notag \\ 
             &= -\kb\int_{\cell_m}\micden( x)\log \left( \vcell_m \micden(x) \right)\df x + \vcell_m \mesden_m \intrv(m)\notag, 
\end{align}
as the {\it fine-grained entropy}.%
In the last line we introduced the quantity
\begin{equation}
 \intrv(m) := \kb\log \left. \vcell_m  \right.,
  \label{eq:IntrinsicEntropyDeterministic}
\end{equation}
which---in analogy to section \ref{sec:stochastic_thermodynamics}---we also call the intrinsic entropy of state $m$.
However, here it is defined in the sense of Boltzmann as the logarithm of a phase space volume rather than in the sense of Gibbs or Shannon.
This identification will be justified also by the final results.

We also introduce the {\it coarse-grained entropy} of cell $m$ using the coarse-grained density $\mesden(x) = \sum_m \mesden_m \chi_{\cell_m}(x) $:
\begin{align}
  \sent{\mesden\mvert\cell_m }  &= -\kb\vcell_m \mesden_m \log\left. {\mesden_m} \right.\\
              &= -\kb\vcell_m \mesden_m \log \left( \vcell_m\mesden_m \right) + \kb\vcell_m \mesden_m \log \left. \vcell_m  \right.\notag\\
              &= -\kb\vcell_m \mesden_m \log \left( \vcell_m\mesden_m \right) + \vcell_m \mesden_m  \intrv(m) \notag\\
              &= \vcell_m\mesden_m(\visrv(m) + \intrv(m)) \notag\\
              &= \vcell_m\mesden_m(\cgrv(m)) \notag 
  \label{eq:cellCGEntropyFull}
\end{align}
The last line splits the microscopic density into two local contributions that are weighted by the factor $\vcell_m \mesden_m$.
The first part is the intrinsic entropy from above, which is a thermodynamic state variable but not directly observable.
The second part is a local quantity, too, but it additionally depends on the ensemble density $\mesden$.
We call this part the apparent or \emph{visible entropy},
\begin{equation}
  \visrv(m) = - \kb \log \left( \vcell_m\mesden_m \right),
  \label{eq:VisibleEntropyDeterministic}
\end{equation}
because this quantity can be inferred from experimentally obtained ensemble statistics.


\subsubsection{Relative information and observable entropy}
Another important quantity is the difference between the coarse- and fine-grained entropy $\sent{\mesden\mvert \cell_m } - \sent{\micden \mvert \cell_m}$, which was  introduced in Ref.~\cite{Breymann_etal1998}  in the context of multi-baker maps.
It can be written as a Kullback-Leibler divergence and is therefore always positive
\begin{equation}
  \sent{\mesden\mvert \cell_m } - \sent{\micden \mvert \cell_m} = \kldiv{\micden \mVert \mesden \mvert \cell_m } = \int_{\cell_m}\micden( x)\log \left. \frac{\micden( x)}{\mesden_m}\right.\mathrm{d} x \label{eq:cellDeltaEntropy}\geq0.
\end{equation}
From information theory we know that this quantity is the relative information of the two descriptions, \ie it quantifies the uncertainty of the microstate of the system if we know the mesostate.
As the system equilibrates from its initially prepared state, this uncertainty becomes larger and therefore this quantity should never decrease.
Later, we will prove this and identify its rate of change with the irreversible, total entropy production rate $\totep$.

%

\subsection{Entropy production and variation}
Under the assumptions of stochastic thermodynamics it is possible to obtain the full-time evolution of the microscopic density, if it was initialized uniformly on the mesoscopic cells.
The idea is to follow the contraction and squeezing of subsets of phase space that move along the same mesoscopic trajectory.

\subsubsection{Discrete trajectories, co-moving microstates and equilibrated cells}
In the stroboscopic picture the evolution operator is a discrete map that generates discrete mesoscopic trajectories.
Therefore the mathematical treatment becomes more lucid and averages and probabilities can be explicitly calculated.
Analogously to the time-continuous phase, we denote a semi-infinite mesoscopic trajectory by $\traj \omega$.
If the trajectory is constrained to run until time $t = \nu\tobs$ it is denoted by $\traj \omega (\nu)$.
It can be represented by a $(\nu + 1)$-tuple:
\begin{equation}
  \traj \omega(\nu) = \left( \omega_0,\omega_1,\ldots,\omega_\nu \right).
  \label{eq:defnDiscreteTraj}
\end{equation}
Where no ambiguity can arise we will write simply $\traj \omega$ instead of $\traj\omega(\nu)$.

Further, we define the mesoscopic history of a point $x \in \cell_{\omega_{\nu}}$,
\begin{equation}
  \traj \omega (x) := \left(\meas(\Phi^{-\nu}x), \meas(\Phi^{-\nu+1}x), \ldots, \meas(\Phi^{-1}x),\meas(x)\right)
  \label{eq:HistoryOfMicrostate}
\end{equation}
and the set of microstates that share the same mesoscopic history $\traj \omega$ 
\begin{equation}
  \tilde\cell\left[\traj \omega\right] := \set{x: \traj\omega(x) = \traj \omega}.
  \label{eq:defnTrajectoryVolumeElement}
\end{equation}
%
%
%
This set can also be recursively defined:
\begin{equation}
  \tilde\cell\left[\traj \omega{(k)}\right] \equiv \Phi\left(\tilde\cell\left[\traj \omega{(k-1)}\right]\cap \cell^{\omega_{k-1}}_{\omega_{k}}  \right).
  \label{eq:defnTrajectoryVolumeElementRecursive}
\end{equation}
The quantity we are interested in is the volume $\vcell\left[\traj \omega\right] := \lambda\left(\tilde\cell\left[\traj \omega\right]  \right)$.
To calculate it, we must formulate the assumption of equilibrated cells for deterministic dynamics.
First recall its meaning in the context of time scales: 
It is the assumption (or better said, approximation) that any set of trajectories entering a cell $\cell_m$ at time $t$ is smeared out over the whole cell after a time-scale $\tau<\tobs$.
Consequently, the set $\tilde\cell\left[\traj \omega^{(k)}\right]$ of microstates that arrived in cell $\cell_{\omega_k}\subset \tilde\cell\left[\traj \omega^{(k)}\right]$ at time $k$ must cover $\cell_{\omega_k}$ densely.
But this is equivalent to saying that the measure of the subset of $\tilde\cell\left[\traj \omega^{(k)}\right]$ that will be mapped to $\cell_{\omega_{k+1}}$ relative to $\vcell\left[\traj\omega^{(k)}\right]$ must be the overall fraction $s^{\omega_k}_{\omega_{k+1}}$:
\begin{equation}
  \frac{\lambda\left( \tilde\cell\left[ \traj \omega^{(k)} \right]\cap\cell^{\omega_{k}}_{\omega_{k=1}} \right)}{\lambda\left(\tilde\cell\left[\traj \omega^{(k)}\right]\right) } \stackrel{!}{=}s^{\omega_k}_{\omega_{k+1}}.
  \label{eq:VolumeFactorizes}
\end{equation}
Note that this is just the Markov assumption, if we interpret $\vcell\left[\traj \omega^{(k)}\right]$ and $s^{\omega_k}_{\omega_{k+1}}$ as the probabilities of observing the trajectory $\traj \omega^{(k)}$ and a jump from $\cell_{\omega_k}$ to $\cell_{\omega_{k+1}}$, respectively:
\begin{equation}
  \text{Prob}\left[\traj\omega^{(k+1)} \right] = \text{Prob}\left[\traj\omega^{(k+1)} \right]\times\text{ Prob}\left[\omega_{k+1}\mvert\omega_{k}\right]
  \label{eq:MarkovAssumption}
\end{equation}

Further, define the contraction factor
\begin{equation}
  \eta^{\omega_k}_{\omega_{k+1}} := \frac{\tilde\vcell^{\omega_{k+1}}_{\omega_{k}}}{ \vcell^{\omega_{k+1}}_{\omega_{k}} } \equiv \frac{\vcell_{\omega_{k+1}}\tilde s^{\omega_{k+1}}_{\omega_{k}}}{\vcell_{\omega_k} s^{\omega_{k+1}}_{\omega_{k}} }.
  \label{eq:contractionFactor}
\end{equation}
Because the sets of co-moving trajectories are smeared out over the whole cell, we assume uniform contraction, \ie
\begin{equation}
  \lambda\left(\Phi\left(\tilde\cell\left[\traj \omega^{(k-1)}\right]\cap \cell^{\omega_{k-1}}_{\omega_{k}}  \right)\right) = \eta^{\omega_{k-1}}_{\omega_j} \lambda\left( \tilde\cell\left[\traj \omega^{(k-1)}\right]\cap \cell^{\omega_{k-1}}_{\omega_{k}}   \right).
  \label{eq:uniformContraction}
\end{equation}

Using equations (\ref{eq:defnTrajectoryVolumeElementRecursive}), (\ref{eq:uniformContraction}), (\ref{eq:VolumeFactorizes}) and the definition (\ref{eq:contractionFactor}) we find a recursive relation for $\vcell\left[\traj\omega\right]$:
\begin{equation}
  \vcell\left[\traj \omega^{(k)}\right] = \frac{\vcell_{\omega_k} }{\vcell_{\omega_{k-1}}}\tilde s^{\omega_{k-1}}_{\omega_k}\vcell\left[\traj\omega^{(k-1)}\right].
  \label{eq:volumeRecursive}
\end{equation}
After iteration we arrive at
\begin{align}
  \vcell\left[\traj \omega^{(\nu)}\right] &= \prod_{k=1}^\nu\left[\frac{\vcell_{\omega_k} }{\vcell_{\omega_{k-1}}}\tilde s^{\omega_{k-1}}_{\omega_k}\right]\vcell\left[\traj\omega^{(k-1)}\right]\nonumber\\
    &=\frac{\vcell_{\omega_\nu} }{\vcell_{\omega_{0}}} \prod_{k=1}^\nu\left[\tilde s^{\omega_{k-1}}_{\omega_k}\right]\vcell_{\omega_0}\nonumber \\
    &=\vcell_{\omega_\nu}  \prod_{k=1}^\nu\left.\tilde s^{\omega_{k-1}}_{\omega_k}\right.\\
    &=\vcell_{\omega_\nu}  \prod_{k=1}^\nu\left.s_{\omega_{k-1}}^{\omega_k}\right.,
  \label{eq:TrajectoryVolumeFinal}
\end{align}
where in the last line we used reversibility.

Here we stress again, that for any real dynamics, the above assumptions have to be understood as an approximation in the same spirit as the Markovian approximation of the stochastic approach.
However, dynamics that exactly fulfill \eq{TrajectoryVolumeFinal} can be constructed.
The most well-known thereof are the reversible, ``properly thermostatted'' Multi-Baker maps used by Vollmer and co-workers \cite{Breymann_etal1998,Vollmer_etal1998,Vollmer2002}.
In that particular case, there is no separation of time-scales or any notion of ``equilibrated cells''.
The subtle (geometrical) reason why one can use \eq{TrajectoryVolumeFinal} is  that the images and pre-images of the map intersect in angles of $90^\circ$.
Therefore, Multi-Baker type maps might be useful as models to study this connection.
However, they do not represent (small) thermodynamic systems.

\subsubsection{Evolution of the densities}

On its way to cell $\omega_\nu$ along trajectory $\traj \omega$ the volume element $\cell\left[\traj\omega\right]$ has been repeatedly stretched and squeezed.
Denote by $\mesden^{\left( k \right)}\left[\traj\omega\right]$ the density of the measure $\mu^{(k)}$ in $\cell\left[\traj\omega\right]$.
Because we start with uniform densities on the cells at time $\nu =0$, this is a constant.
Probability conservation implies that
\begin{equation}
  \mesden^{(k)}\left[\traj \omega\right] \vcell_{\omega_k}s^{\omega_{k}}_{\omega_{k+1}} = \mesden^{(k+1)}\left[\traj \omega\right] \vcell_{\omega_{k+1}}s^{\omega_{k+1}}_{\omega_k}.
  \label{eq:evolutionOfLocalDensities}
\end{equation}
This leads to
\begin{equation}
  \mesden^{(k+1)}\left[\traj \omega\right] = \mesden^{(k)}\left[\traj \omega\right]\frac{\vcell_{\omega_k}s^{\omega_{k}}_{\omega_{k+1}} }{\vcell_{\omega_{k+1}}s^{\omega_{k+1}}_{\omega_k}}.
\end{equation}
After iteration we find the density at a point $x \in \vcell_{\omega_\nu}$ with a trajectory $\traj \omega(x)$:
\begin{equation}
  \micden^{(\nu)}(x) = \mesden^{(\nu)}\left[\traj \omega(x)\right] = \mesden_{\omega_0} \prod_{k=1}^{\nu}\left[\frac{s^{\omega_{k-1}}_{\omega_k}}{s^{\omega_{k}}_{\omega_{k-1}}} \right]\frac{\vcell_{\omega_0}}{\vcell_{\omega_\nu}}.
  \label{eq:evolutionMicroscopicDensities}
\end{equation}
The evolution of the coarse-grained, mesoscopic density $\rho^{(\nu)}$ is obtained by integration over cell $\vcell_{\omega_{\nu}}$:
\begin{equation}
  \vcell_{\omega_{\nu}}\mesden_{\omega_\nu}^{(\nu)} = \left. \sum_{\traj\omega}\right.^{\omega_\nu}\left[ \mesden_{\omega_0}\vcell_{\omega_0} \prod_{k=1}^\nu s^{\omega_{k-1}}_{\omega_{k}} \right]
  \label{eq:mesdenSolution}
\end{equation}
Here, $\sum_{\traj \omega}^{m}$ means that we sum over all trajectories that end in cell $\cell_m$.
The weight $p_m$ of cell $\cell_m$ with respect to the measure $\mu^{\left( k \right)}$ is
\begin{equation}
  p_m^{(k)}:= \mu^{\left( k \right)}(\cell_m) = \vcell_m\mesden_m^{(k)}.
  \label{eq:defnProbabilityDeterministic}
\end{equation}
It is easy to see that \eq{mesdenSolution} is the solution of the discrete-time Master equation
\begin{equation}
  p_i^{(k+1)} = \sum_j\left[ s^j_i p_j^{(k)}\right].
  \label{eq:DiscreteTimeME}
\end{equation}
under the assumptions that we found an initially uniform density on each cell, \ie $p^{(0)}_{m} = \vcell_m \mesden_m$.
In \eq{DiscreteTimeME} the $s^i_j$ take the role of transition probabilities.
Now we define the weight of a mesoscopic trajectory $\traj \omega$
\begin{equation}
  \mathbb P[\traj \omega] = p_{\omega_0}^{(0)} \prod_{k=1}^\nu s^{\omega_{k-1}}_{\omega_{k}} .
  \label{eq:TrajectoryWeight}
\end{equation}
It obeys 
\begin{align}
  \sum_{\traj \omega} \mathbb P \left[\traj \omega\right]&=1,\\
  \left.\sum_{\traj \omega}\right.^{\omega_\nu}\mathbb P \left[\traj \omega\right] &= p^{(\nu)}_{\omega_{\nu}}.
\end{align}
Inserting \eq{evolutionMicroscopicDensities} into the definition of the fine-grained entropy \eq{cellFGEntropy} we find
\begin{align}
  \sent{\micden^{(\nu)}\mvert \cell_{\omega_\nu}} &=
  -\kb \left.\sum_{\traj \omega}\right.^{\omega_\nu} \left[\vcell_{\omega_{0}}\mesden_{\omega_0}  \prod_{k=1}^\nu\left[ s^{\omega_{k-1}}_{\omega_k} \right] \left( \sum_{k=1}^{\nu}\left[ \log\frac{s^{\omega_{k-1}}_{\omega_{k}}}{s^{\omega_{k}}_{\omega_{k-1}}} \right]
  + \log \left(\vcell_{\omega_0}\mesden_{\omega_{0}}\right) \right)\right] \nonumber\\
  &\quad+ \vcell_{\omega_{\nu}}\mesden^{(\nu)}_{\omega_{\nu}}\intrv(\omega_{\nu}).
  \label{eq:LocalFGEntropy}
\end{align}
Also here we define $B^{i}_j := \kb\log\left.\frac{s^i_j}{s^j_i}\right.$ using the volume fractions $s^i_j$.
In this context it can be interpreted as the rate of phase space contraction \cite{Vollmer2002}.
A more suitable form is:
\begin{align}
  \sent{\micden^{(\nu)}\mvert \cell_{\omega_\nu}}  &=- \left.\sum_{\traj \omega}\right.^{\omega_\nu} \left[\mathbb P[\traj \omega]  \left. \sum_{k=1}^{\nu} B^{\omega_{k-1}}_{\omega_{k}}\right. \right] \nonumber \\ & \quad - \kb\sum_{\omega_0}\left[p^{(0)}_{\omega_{0}}\log\left. p^{(0)}_{\omega_{0}} \right]\right. +  p^{(\nu)}_{\omega_{\nu}}\intrv(\omega_{\nu}).
  \label{eq:LocalFGEntropyRewritten}
\end{align}
The first term is the entropy originating from the jumps of the trajectories that end in $\vcell_{\omega_\nu}$ at time $\nu$.
The second term is the visible entropy specified by the initial conditions.
The third term is the part of the intrinsic entropy associated with cell $\omega_\nu$.

\subsubsection{Recovering the trajectory functionals of ST}
At this time it is worthwhile to stop and reflect whether it makes sense to attempt an interpretation of \eq{LocalFGEntropyRewritten} as the ``fine-grained entropy of cell $\omega_{\nu}$ at time $\nu$''.
The problem is that such an interpretation is local in space and time.
However, even though we could eliminate the microscopic density, the first term now depends on the history of many (non-local) mesoscopic trajectories.
Therefore it is impossible to describe this term without referring to the other cells or the memory of the process.

Let us recall the functionals (\ref{eq:IntrinsicEntropyTraj}--\ref{eq:JumpEntropy}) from section \ref{sec:stochastic_thermodynamics}.
With the modified definitions of the intrinsic entropy, the  weight of a cell $p_m$ and the contraction rate $B^i_j$, the discrete analogues look formally the same:
\begin{align}
  \intrv[\traj \omega,\nu] &= \intrv\left(\omega_\nu\right) = \kb \log \vcell_{\omega_\nu},\label{eq:IntrinsicEntropyDisc}\\
  \visrv[\traj \omega,\nu] &= \visrv(\omega_\nu)= - \kb \log p_{\omega_\nu}^{(\nu)},\label{eq:VisibleEntropyDisc}  \\
  \jumprv[\traj \omega,\nu] &= \sum_{j=1}^{\nu} B^{\omega_{j-1}}_{\omega_j} \label{eq:JumpEntropyDisc}
\end{align}
With this we can express the sum of the fine-grained entropy of all cells with the help of trajectory averages
\begin{align}
  S^{(\nu)}_{\mathrm{fg}} &:= \sum_{\omega_\nu} \sent{\micden^{(\nu)}\mvert \cell_{\omega_\nu}} \label{eq:fgEntropy}\\
  &=  - \left.\sum_{\traj \omega}\right. \left[\mathbb P[\traj \omega]  \left. \sum_{k=1}^{\nu} B^{\omega_{k-1}}_{\omega_{k}}\right. \right] \nonumber \\
  & \quad - \kb\sum_{\omega_0}\left[p^{(0)}_{\omega_{0}}\log\left. p^{(0)}_{\omega_{0}} \right.\right] +  \sum_{\omega_\nu} \left[p^{(\nu)}_{\omega_{\nu}}\intrv(\omega_{\nu})\right]\\
  &= -\trav{ \jumprv } + \eav{\visrv}^{\left( 0 \right)} + \trav{ \intrv }
  \label{eq:FGEntropyRewritten}
\end{align}
The first and the last term of the last line are trajectory averages of the form of \eq{TrajectoryAverageGeneral} though the latter could also be expressed by a trajectory average of a local quantity using \eq{AverageOfLocalObservable}.
The second term is the visible entropy of the initial ensemble.

The coarse-grained density of all cells reads:
\begin{align}
  S^{(\nu)}_{\mathrm{cg}} &:= \sum_{\omega_\nu} \sent{\mesden^{(\nu)}\mvert \cell_{\omega_\nu}}\label{eq:cgEntropy} \\
   &=- \kb\sum_{\omega_\nu}\left[p^{(\nu)}_{\omega_{\nu}}\log\left. p^{(\nu)}_{\omega_{\nu}} \right.\right] +  \sum_{\omega_\nu} \left[p^{(\nu)}_{\omega_{\nu}}\intrv(\omega_{\nu})\right] \label{eq:splitCGentropy} \\
  &= \eav {\visrv}^{(\nu)} + \eav{\intrv}^{\left( \nu \right)} \\
  &= \trav {\visrv} + \trav{\intrv} 
  \label{eq:CGEntropyRewritten}
\end{align}
For the difference entropy $\klent$ we find
\begin{equation}
  \klent^{(\nu)} = \eav {\visrv}^{(\nu)} -\eav {\visrv}^{(0)} + \trav \jumprv  = \visent^{(\nu)}-\visent^{(0)} + \trav\jumprv.
  \label{eq:differenceEntropyRewritten}
\end{equation}

The main result is that these identifications allow us to relate the physical entropies to the (Shannon-)entropies in the context of deterministic dynamics:
\begin{align}
  \sysent^{(\nu)} = \trav \sysrv  &\equiv  \cgent^{(\nu)}, \\
  \medent^{(\nu)} = \trav \medrv  &\equiv -\fgent^{(\nu)} + \visent^{(0)},\\
  \totent^{(\nu)} = \trav \totrv &\equiv \klent^{(\nu)} - \visent^{(0)}.
\end{align}
Because of the one-to-one correspondence, the relation to the three fundamental functionals (\ref{eq:SystemEntropyTraj}-\ref{eq:TotalEntropyTraj}) hold as well.

It remains to find the analogues of the entropy variations $\sigma$.
In the discrete-time case, this is easy, as $\sigma[\traj \omega,\nu]$ can be obtained as $s[\traj\omega,\nu] - s[\traj\omega,\nu-1]$ for any discrete entropy functional $s$:
\begin{eqnarray}
  \viseprv[\traj \omega,\nu] &=& \kb\log\left. \frac{p_{\omega_\nu}^{(\nu)}}{p_{\omega_{\nu-1}}^{(\nu-1)}}\right.,\label{eq:visEntropyRateDisc} \\
  \inteprv[\traj\omega,\nu] &=&  \intrv(\omega_\nu) - \intrv(\omega_{\nu-1}),\\
  \jumpeprv[\traj\omega,\nu] &=& B^{\omega_{\nu-1}}_{\omega_\nu} \label{eq:BEntropyRateDisc}. 
\end{eqnarray}

Again, from these quantities one can construct the functionals of the physical entropy variations per unit time as well as their trajectory averages.
A surprise in the discrete case may be the result for the average total entropy production per unit time $\totep = \trav{\viseprv + \jumpeprv}$:
\begin{align}
    \totep = & \sum_{i,j}\left[ p_i s^i_j B^i_j \right] - \kb\sum_i\left[ p_i'\log p_i' - p_i\log p_i \right]\notag\\
    =& \kb\sum_{i,j}\left[ p_i s^i_j \log \frac{p_is^i_j}{p_j's^j_i} \right] \notag\\
    =& \kb\sum_{i,j}\left[ p_i s^i_j \log \frac{p_is^i_j}{p_js^j_i} \right] + \kb\sum_i\left[ p_i'\log \frac{p_i}{p_i'} \right]  >0.
  \label{eq:finiteTimeTotEP}
\end{align}
The average total entropy production is, as expected, always positive.
This can be seen from the second line where we write it as a Kullback-Leibler divergence.
However, looking at the third line we see that it differs from the direct correspondence to the continuous case by a boundary term.
Also these two terms have the form of Kullback-Leibler divergences.
The negative second term becomes increasingly smaller as a steady state is reached.
It has often been ignored by previous works because it does not appear neither in the continuous case nor in the steady state.

In summary, we constructed the discrete versions of the fundamental functionals used in stochastic thermodynamics from deterministic dynamics.

\subsection{Connection to the continuous time description}
To arrive at the discrete-time analogues of the trajectory functionals one basically just has to substitute $w^i_j$ by $s^i_j$ in the continuous-time expressions (\ref{eq:VisibleEntropyTraj}--\ref{eq:JumpEntropyRateTraj}).
The variations $\sigma$ were obtained as differences of the entropies.
However, they can also be obtained from the continuous variations $\sigma^c$, which we will denote by the superscript $c$, if we consider a given continuous trajectory $\traj \omega$:

Let us consider the time interval $[\nu\tobs,(\nu+1)\tobs]$.
We assume that $\tobs$ is small enough that at most one jump occurs at $\tau_j$ from states $\omega_- := \omega\left((\nu-1)\tobs\right)\equiv \omega_{\nu-1}$ to $\omega_+ :=\omega\left(\nu\tobs\right) \equiv \omega_{\nu} $.
An entropy functional $s$ can be written as the integral over the continuous-time derivative functional:
\begin{equation}
  \begin{split}
  \sigma = & s[\traj \omega; (\nu+1) \tobs] - s[\traj\omega, \nu\tobs]\\
  = & \int_{\nu\tobs}^{(\nu+1)\tobs} \sigma^c[\traj\omega(t)] \df t\\
  = & \int_{\nu\tobs}^{\tau_j} \sigma^c[\traj\omega(t)] \df t +  \int_{\tau_j}^{(\nu+1)\tobs} \sigma^c[\traj \omega(t)] \df t
  \end{split}
  \label{eq:GeneralIntegrationDifferenceFunctional}
\end{equation}
In the last line we split the integral into two contributions.
This is important in order to evaluate the ensemble probability for the correct states.
The first line implies that in the discrete case $\Sigma$ is the average variation per unit time of the corresponding average value of $S$.

For completeness, we mention that the continuous time description is usually obtained in a more formal way.
This is done by converting the expressions in the discrete-time Master equation (\ref{eq:DiscreteTimeME}) into its continuous-time analogue, \eq{MasterEquation}, in a limit procedure $\tobs \to 0$.
The jump probabilities $s^i_j$ for time spans $\tobs$ are interpreted as generated by the infinitesimal generator $(W)_{ij} = (w^i_j)$ \cite{Feller1968}:
\begin{equation}
  \begin{split}
  s^i_j =& \exp\left( w^i_j\tobs \right),\quad (i\neq j),\\
  w^i_i =& -\sum_j w^i_j.
  \end{split}
  \label{eq:ratesFromProbabilites}
\end{equation}

\subsection{Connection to previous works}
For consistency, we partly introduced new notation for different entropic quantities.
Random variables are always denoted by minuscules, with their (trajectory) averages represented by the corresponding upper case letters.
Further $\Sigma$ denotes the time derivative or variation per unit time of its entropy $S$, in the continuous or discrete case, respectively.

As a summary, we discuss the main entropic terms identified in this work again in detail and  give an (incomplete) overview of the notation and interpretation in previous work.

\paragraph{Visible entropy}
The visible entropy \visent was defined in \eq{ShannonEntOfEnsemble} as the Shannon entropy of the (discrete) ensemble.
In the same spirit it is the observable part of the coarse-grained entropy, \cf \eq{splitCGentropy}. It can also be expressed as the average over the entropy functional $\visrv$.

It has been often simply called entropy $S$ (in the context of Markov chains) \cite{Schnakenberg1976,Gaspard2004,Andrieux+Gaspard2007,Esposito+vdBroeck2010}.
However, Lebowitz and Spohn have called it also Gibbs entropy $S_G$ because of its functional form and to distinguish it from Boltzmann entropy.

\paragraph{Intrinsic entropy}
The intrinsic entropy $\intrv(\omega)$ is the unobservable (local) equilibrium entropy of a mesoscopic state $\omega$.
It can be calculated canonically from the constrained microscopic density as in \eq{IntrinsicEntropyStochastic}.
On the phase space of the thermostatted, deterministic system it is calculated in the sense of Boltzmann as the logarithm of the volume of a cell (\ref{eq:IntrinsicEntropyDeterministic}) \cite{Seifert2011,Seifert2012}.
In this spirit, is has also been called Boltzmann entropy $S_B(M)$ of (mesoscopic) state $M$ \cite{Maes+Netocny2003}.

\paragraph{Fine-grained entropy}
The coarse-grained entropy \fgent is the Shannon entropy (\ref{eq:fgEntropy}) of the microscopic (Liouville) density $\micden$.
For equilibrium systems, it is equivalent to Gibbs' entropy.
Maes and Neto{\v{c}}n{\'y} denote it $S(\mu)$ and make a strong point for not calling it Gibbs entropy \cite{Maes+Netocny2003}, which is the expression used by Vollmer and co-workers, who denoted it $S^{(G)}$ \cite{Breymann_etal1998,Vollmer2002}.

\paragraph{Coarse-grained entropy}
The coarse-grained entropy \cgent is the Shannon entropy of the coarse-grained density $\mesden$ from \eq{cgEntropy}.
It consists of the visible entropy \visent and the intrinsic entropy of the current ensemble $\intent = \eav{\intrv}$.
Vollmer and co-workers call it simply the entropy $S$ \cite{Breymann_etal1998,Vollmer2002}, whereas Maes and Neto{\v{c}}n{\'y} denote it the Shannon entropy of the (projected, \ie coarse-grained) measure, $\hat S(\hat \mu)$ \cite{Maes+Netocny2003}.

\paragraph{Total entropy production}
The total positive entropy production \totep is the quantity to be identified with the total dissipation \cite{Seifert2005,Seifert2012}.
It is always positive due to its form of a Kullback-Leibler divergence.
In the context of information theory it is the rate of increase of the uncertainty of the microscopic configuration over time, if one keeps track of the mesoscopic densities.

It has previously also been called the internal non-negative entropy production \cite{Schnakenberg1976,Gaspard2004,Andrieux+Gaspard2007,Esposito+vdBroeck2010}.
Notation has varied a lot including the symbols $P$ \cite{Schnakenberg1976}, $R$ \cite{Lebowitz+Spohn1999},  $\frac 1 \tau \Delta_i^\tau$ \cite{Gaspard2004}, $\tdiff{_iS}{t}$ \cite{Andrieux+Gaspard2007}, $\dot S_i$ \cite{Esposito+vdBroeck2010} and $\dot{S}^{\mathrm{tot}}$ \cite{Seifert2005}.
Usually, it has been considered for continuous-time Markov chains where it has the form of the first term in \eq{functionalsToSchnakenberg}.
However, as we saw above, for the discrete situation an additional boundary term from the finite time interval appears with respect to the continuous-time formulation.
This was already noted by Gaspard \cite{Gaspard2004}, though his identification and interpretation is different than in the present context.

\paragraph{Motance entropy variation}
The variation of the motance entropy \jumpep is facilitated by the local motance $B^i_j$.
As we see from the deterministic discussion (\ref{eq:FGEntropyRewritten}, the motance \jumpent is the main part of the fine-grained entropy and directly related to phase space contraction.
Thermodynamically, it is the heat released to the medium less the variation of the intrinsic entropy \cite{Seifert2011,Seifert2012}.

The latter contribution has often been neglected when \jumpep has meed identified with the entropy change in the reservoirs describing the medium of the environment \cite{Schnakenberg1976,Esposito+vdBroeck2010}.
In the same sense it is often called the entropy flux to the medium or external entropy variation \cite{Gaspard2004,Andrieux+Gaspard2007}.
As for the case of the total entropy production, symbols used include $P_2$ \cite{Schnakenberg1976}, $A$ \cite{Lebowitz+Spohn1999},  $-\frac 1 \tau \Delta_e^\tau$ \cite{Gaspard2004}, $-\tdiff{_eS}{t}$ \cite{Andrieux+Gaspard2007}, $-\dot S_e\equiv -\dot S_r$ \cite{Esposito+vdBroeck2010} and $\dot{S}^{\mathrm{tot}}$ \cite{Seifert2005}.

\subsection{Application: Isolated systems}
As Seifert stresses \cite{Seifert2011}, it is crucial that the motance consists of the variation of intrinsic entropy as well as the entropy flux to the medium.
Here, we show this explicitly using an isolated system evolving under Hamiltonian dynamics.
We know that the invariant measure $\micden^\infty$ will be uniform on the constrained phase space (energy shell), corresponding to an equilibrium situation.
Further, detailed balance should hold between mesoscopic cells $\lambda(\cell_i) = \vcell_i$.
Without loss of generality we set the total phase space volume to unity, \ie $\lambda(\pspace) = \sum_i \vcell_i = 1$.
Then, because $\micden^\infty =1$, we find that the equilibrium probabilities obey $p^\infty_i = \vcell_i$.
From the detailed balance condition
\begin{equation}
  p^\infty_is^i_j = p^\infty_j s^j_i
  \label{eq:detailedBalance}
\end{equation}
we then infer that $s^i_j = c_{ij}\vcell_j$ with some constant $c_{ij}=c_{ji}$.
But because $\sum_j s^i_j =1$ for all $i$, one has to demand $c_{ji}=1$, $\forall i,j$.

By definition, an isolated system cannot allow any ``entropy flux to the medium'' at any time, \ie
\begin{equation}
  \trav \medeprv = \trav \jumpeprv -\trav \inteprv \stackrel{!}{=} 0.
  \label{eq:isolatedSystem}
\end{equation}
But this can be seen from
\begin{align*}
  \trav \medeprv &= \sum_{i,j}\left[ p_is^i_j\log \frac{s^i_j}{s^j_i} \right]-\sum_{i}\left[ \left( p'_i - p_i \right)\log\vcell_i \right]\\
  &= \sum_{i,j}\left[ p_i\vcell_j \log  \vcell_j -p_i\vcell_j\log \vcell_i -p_j\vcell_i\log \vcell_i \right]  \\ 
      &\quad +\sum_i\left[ p_i\log \vcell_i \right] \\
  &= \sum_{i,j}\left[ -p_i\vcell_j \log \vcell_i \right]+\sum_i\left[ p_i\log \vcell_i  \right]\\
  &= \sum_i\left[ \left( -p_i + p_i)\log \vcell_i  \right) \right]= 0.
\end{align*}

\section{Conclusion}
In this work we presented a new approach to the entropy functionals of stochastic thermodynamics using deterministic evolution equations.
We used reversible deterministic thermostatted equations of motion as a paradigm.
The finite resolution of any real measurement motivated the introduction of time-reversal invariant mesoscopic observables and the use of a stroboscopic picture.
Under the assumptions of equilibrated cells we could follow the evolution of an observable coarse-grained and an unobservable fine-grained density.
This lead to the notion of coarse- and fine-grained entropy, whose dynamics is intimately related to the notion of entropy and entropy production in the system and its environment.
Finally, we were able to construct the trajectory-dependent functionals of stochastic thermodynamics from a purely deterministic dynamics.
We also made the connection of the discrete-time case to the more common continuous-time description.

This work also hopes to disentangle some of the confusion in the context of nonequilibrium entropies.
Maybe it can contribute to reconcile different views on entropy and entropy production through the one-to-one correspondence of phase space contraction and thermodynamic interpretation of (ratios) of transition probabilities.

Further, the special role of Multi-Baker-Maps in the context of nonequilibrium thermodynamics was analysed:
Rather than as simplified models for chaotic, many-particle dynamical systems they have to be understood as mathematical models that show a similar evolution, due to a geometric peculiarity.
However, they might elucidate the connection between dynamical systems and stochastic thermodynamics further and future work in this direction is in progress.

\ack
The author wants to thank Lamberto Rondoni, J\"urgen Vollmer, Thomas Frerix and Artur Wachtel for many fruitful discussions.
Further, I am indebted to Artur Wachtel and Lukas Geyrhofer for comments on the manuscript.

\clearpage

\section*{References}
\bibliographystyle{plain}
\bibliography{/home/baltaner/bib/noneq/noneq,/home/baltaner/bib/misc/misc}

\begin{thebibliography}{10}

\bibitem{Altaner_etal2012}
B.~Altaner, S.~Grosskinsky, S.~Herminghaus, L.~Katth{\"a}n, M.~Timme, and
  J.~Vollmer.
\newblock Network representations of nonequilibrium steady states: Cycle
  decompositions, symmetries, and dominant paths.
\newblock {\em Phys. Rev. E.}, 85(4):041133, 2012.

\bibitem{Altaner+Vollmer2012}
Bernhard Altaner and J\"urgen Vollmer.
\newblock Fluctuation-preserving coarse graining for biochemical systems.
\newblock {\em Phys. Rev. Lett.}, 108:228101, 2012.

\bibitem{Andrieux+Gaspard2007}
D.~Andrieux and P.~Gaspard.
\newblock Fluctuation theorem for currents and {S}chnakenberg network theory.
\newblock {\em J. Stat. Phys.}, 127(1):107--131, 2007.

\bibitem{Braga+Travis2005}
C.~Braga and K.P. Travis.
\newblock A configurational temperature nos{\'e}-hoover thermostat.
\newblock {\em J. Chem. Phys.}, 123:134101, 2005.

\bibitem{Breymann_etal1998}
W.~Breymann, T.~T{\'e}l, and J.~Vollmer.
\newblock Entropy balance, time reversibility, and mass transport in dynamical
  systems.
\newblock {\em Chaos}, 8(2):396--408, 1998.

\bibitem{Crooks1999}
G.E. Crooks.
\newblock Entropy production fluctuation theorem and the nonequilibrium work
  relation for free energy differences.
\newblock {\em Phys. Rev. E.}, 60(3):2721, 1999.

\bibitem{deGroot+Mazur1984}
S.R. de~Groot and P.~Mazur.
\newblock {\em Non-equilibrium thermodynamics}.
\newblock Dover, 1984.

\bibitem{Esposito2012}
M.~Esposito.
\newblock Stochastic thermodynamics under coarse graining.
\newblock {\em Phys. Rev. E.}, 85(4):041125, 2012.

\bibitem{Esposito+vdBroeck2010}
M.~Esposito and C.~Van~den Broeck.
\newblock Three faces of the second law. {I.} {M}aster equation formulation.
\newblock {\em Phys. Rev. E}, 82(1):011143, Jul 2010.

\bibitem{Evans_etal1993}
D.~J. Evans, E.~G.~D. Cohen, and G.~P. Morriss.
\newblock Probability of second law violations in shearing steady states.
\newblock {\em Phys. Rev. Lett.}, 71:2401--2404, Oct 1993.

\bibitem{Faggionato+dPietro2011}
A.~Faggionato and D.~Di~Pietro.
\newblock {G}allavotti--{C}ohen-type symmetry related to cycle decompositions
  for {M}arkov chains and biochemical applications.
\newblock {\em J. Stat. Phys.}, 143:11--32, 2011.

\bibitem{Falcioni_etal2007}
M.~Falcioni, L.~Palatella, S.~Pigolotti, L.~Rondoni, and A.~Vulpiani.
\newblock Initial growth of boltzmann entropy and chaos in a large assembly of
  weakly interacting systems.
\newblock {\em Physica A}, 385(1):170--184, 2007.

\bibitem{Feller1968}
W.~Feller.
\newblock {\em {An Introduction to Probability Theory and Its Applications}},
  volume~1.
\newblock Wiley, 3 edition, 1968.

\bibitem{Gallavotti+Cohen1995}
G.~Gallavotti and E.~Cohen.
\newblock Dynamical ensembles in stationary states.
\newblock {\em J. Stat. Phys.}, 80:931--970, 1995.

\bibitem{Gardiner2004}
C.W. Gardiner.
\newblock {\em Stochastic methods}.
\newblock Springer Berlin, 4th edition, 2004.

\bibitem{Gaspard2004}
P.~Gaspard.
\newblock Time-reversed dynamical entropy and irreversibility in markovian
  random processes.
\newblock {\em J. Stat. Phys.}, 117:599--615, 2004.

\bibitem{Hill1977}
T.L. Hill.
\newblock {\em Free energy transduction in biology}.
\newblock Academic Press New York, 1977.

\bibitem{Jarzynski1997}
C.~Jarzynski.
\newblock Nonequilibrium equality for free energy differences.
\newblock {\em Phys. Rev. Lett.}, 78(14):2690, 1997.

\bibitem{Jaynes1957}
E.~T. Jaynes.
\newblock Information theory and statistical mechanics.
\newblock {\em Phys. Rev.}, 106:620--630, May 1957.

\bibitem{Jepps+Rondoni2010}
O.G. Jepps and L.~Rondoni.
\newblock Deterministic thermostats, theories of nonequilibrium systems and
  parallels with the ergodic condition.
\newblock {\em J. Phys. A}, 43:133001, 2010.

\bibitem{Kurchan1998}
J.~Kurchan.
\newblock Fluctuation theorem for stochastic dynamics.
\newblock {\em J. Phys. A}, 31:3719, 1998.

\bibitem{Lebowitz+Spohn1999}
J.L. Lebowitz and H.~Spohn.
\newblock A {G}allavotti--{C}ohen type symmetry in the large deviation
  functional for stochastic dynamics.
\newblock {\em J. Stat. Phys.}, 95:333, 1999.

\bibitem{Maes2004}
C.~Maes.
\newblock {On the origin and the use of fluctuation relations for the entropy}.
\newblock In {\em Poincar{\'e} Seminar 2003: Bose-Einstein
  condensation-entropy}, page 145. Birkh{\"a}user, Basel, 2004.

\bibitem{Maes+Netocny2003}
C.~Maes and K.~{Neto{\v{c}}n{\'y}}.
\newblock Time-reversal and entropy.
\newblock {\em J. Stat. Phys.}, 110(1):269--310, 2003.

\bibitem{Mandal+Jarzynski2012}
D.~Mandal and C.~Jarzynski.
\newblock Work and information processing in a solvable model of maxwell's
  demon.
\newblock {\em P.N.A.S.}, 109(29):11641--11645, 2012.

\bibitem{Ruelle1999}
D.~Ruelle.
\newblock Smooth dynamics and new theoretical ideas in nonequilibrium
  statistical mechanics.
\newblock {\em J. Stat. Phys.}, 95(1):393--468, 1999.

\bibitem{Rugh1997}
H.H. Rugh.
\newblock Dynamical approach to temperature.
\newblock {\em Phys. Rev. Lett.}, 78(5):772, 1997.

\bibitem{Sagawa+Ueda2010}
T.~Sagawa and M.~Ueda.
\newblock Generalized jarzynski equality under nonequilibrium feedback control.
\newblock {\em Phys. Rev. Lett.}, 104(9):90602, 2010.

\bibitem{Samoletov_etal2010}
AA~Samoletov, CP~Dettmann, and MA~Chaplain.
\newblock Notes on configurational thermostat schemes.
\newblock {\em J. Chem. Phys.}, 132(24):246101, 2010.

\bibitem{Samoletov_etal2007}
A.A. Samoletov, C.P. Dettmann, and M.A.J. Chaplain.
\newblock Thermostats for “slow” configurational modes.
\newblock {\em J. Stat. Phys.}, 128(6):1321--1336, 2007.

\bibitem{Schnakenberg1976}
J.~Schnakenberg.
\newblock {Network theory of microscopic and macroscopic behavior of master
  equation systems}.
\newblock {\em Rev. Mod. Phys.}, 48(4):571--585, 1976.

\bibitem{Schulman2005}
L.S. Schulman.
\newblock {\em Techniques and applications of path integration}.
\newblock Dover Publications, 2005.

\bibitem{Seifert2005}
U.~Seifert.
\newblock {Entropy production along a stochastic trajectory and an integral
  fluctuation theorem}.
\newblock {\em Phys. Rev. Lett.}, 95(4):40602, 2005.

\bibitem{Seifert2011}
U.~Seifert.
\newblock Stochastic thermodynamics of single enzymes and molecular motors.
\newblock {\em Eur. Phys. J. E}, 34(3):1--11, 2011.

\bibitem{Seifert2012}
U.~Seifert.
\newblock Stochastic thermodynamics, fluctuation theorems, and molecular
  machines.
\newblock {\em arXiv:1205.4176}, 2012.

\bibitem{Shannon1948}
C.~E. Shannon.
\newblock A mathematical theory of communication.
\newblock {\em Bell System Technical Journal}, 27:379--–423, 623--656, 1948.
\newblock many reprints available.

\bibitem{Verlinde2011}
E.~Verlinde.
\newblock On the origin of gravity and the laws of newton.
\newblock {\em Journal of High Energy Physics}, 2011(4):1--27, 2011.

\bibitem{Vollmer2002}
J.~Vollmer.
\newblock Chaos, spatial extension, transport, and non-equilibrium
  thermodynamics.
\newblock {\em Phys. Rep.}, 372(2):131--267, 2002.

\bibitem{Vollmer_etal1998}
J.~Vollmer, T.~T{\'e}l, and W.~Breymann.
\newblock Entropy balance in the presence of drift and diffusion currents: an
  elementary chaotic map approach.
\newblock {\em Phys. Rev. E.}, 58(2):1672, 1998.

\end{thebibliography}

\end{document}